\documentclass[11pt,a4paper]{emulateapj}
\usepackage{epsfig}
\usepackage{amsmath}
\usepackage{natbib}

\def\spose#1{\hbox to 0pt{#1\hss}}
\def\simlt{\mathrel{\spose{\lower 3pt\hbox{$\mathchar"218$}}
     \raise 2.0pt\hbox{$\mathchar"13C$}}}
\def\simgt{\mathrel{\spose{\lower 3pt\hbox{$\mathchar"218$}}
     \raise 2.0pt\hbox{$\mathchar"13E$}}}

\usepackage{graphicx}

\usepackage{graphicx}

\begin{document}

\title{The Mid-Infrared Environments of High-Redshift Radio Galaxies}

\author{Audrey Galametz\altaffilmark{1,2}, Daniel Stern\altaffilmark{1}, Carlos De Breuck\altaffilmark{3},
Nina Hatch\altaffilmark{4}, Jack Mayo\altaffilmark{5}, \\
George Miley\altaffilmark{6}, Alessandro Rettura\altaffilmark{7},
Nick Seymour\altaffilmark{8}, S. Adam Stanford\altaffilmark{9},
Jo\"{e}l Vernet\altaffilmark{3}}

\altaffiltext{1}{Jet Propulsion Laboratory, California Institute of Technology, 4800 Oak Grove Dr., Pasadena, CA 91109, USA}
\altaffiltext{2}{INAF - Osservatorio di Roma, Via Frascati 33, I-00040, Monteporzio, Italy [e-mail: {\tt audrey.galametz@oa-roma.inaf.it}]}
\altaffiltext{3}{European Southern Observatory, Karl-Schwarzschild-Strasse 2, D-85748 Garching, Germany}
\altaffiltext{4}{University of Nottingham, School of Physics and Astronomy, Nottingham NG7 2RD}
\altaffiltext{5}{Institute for Astronomy, Royal Observatory, Blackford Hill, Edinburgh, EH9 3HJ, UK}
\altaffiltext{6}{Leiden Observatory, University of Leiden, P.B. 9513, Leiden 2300 RA, The Netherlands}
\altaffiltext{7}{Department of Physics and Astronomy, University of California, Riverside, CA 92521, USA} 
\altaffiltext{8}{Mullard Space Science Laboratory, UCL, Holmbury St Mary, Dorking, Surrey, RH5 6NT}
\altaffiltext{9}{Institute of Geophysics and Planetary Physics, Lawrence Livermore National Laboratory, Livermore, CA 94550}

\begin{abstract}

{Taking advantage of the impressive sensitivity of {\it Spitzer}
to detect massive galaxies at high redshift, we study the mid-infrared
environments of powerful, high-redshift radio galaxies at $1.2<z<3$. 
Galaxy cluster member candidates were isolated using a single {\it Spitzer}/IRAC 
mid-infrared color criterion, $[3.6]-[4.5]>-0.1$ (AB), in the fields of $48$ radio galaxies at 
$1.2<z<3$. 
Using a counts-in-cell analysis, we identify a field as overdense when $15$ 
or more red IRAC sources are found within $1\arcmin$ (i.e.,~$0.5$~Mpc at $1.2<z<3$) 
of the radio galaxy to the $5\sigma$ flux density limits of our IRAC data 
($f_{4.5}=13.4\mu$Jy). We find that radio galaxies lie preferentially in medium 
to dense regions, with $73\%$ of the targeted fields denser than average. Our (shallow) 
$120$s data permit the rediscovery of previously known clusters 
and protoclusters associated with radio galaxies as well as the discovery of new promising 
galaxy cluster candidates at $z>1.2$.}

\end{abstract}



\section{Introduction}


The classical technique of finding distant galaxy clusters from the
extended X-ray emission associated with their intracluster medium
is very effective at finding clusters out to $z\sim1$, but rapidly
becomes insensitive for cluster searches beyond $z\sim1.2$ (for a
detailed review of X-ray selected clusters, see Rosati et 
al.~2002)\nocite{Rosati2002}. The challenge of using X-ray observations
to find the most distant clusters is primarily due to the $(1
+ z)^4$ fading of the X-ray surface brightness.
While X-ray selection is very effective at finding massive structures
at moderate redshifts and large samples are expected from the all-sky
eROSITA soft X-ray telescope 
after it launches in 2013 \citep{Cappelluti2011}, X-ray selection
is unlikely to find the most distant clusters and proto-clusters.

Another proven method to identify distant clusters is by searching
wide-field imaging surveys for the red sequence of early-type
galaxies. Such red sequence searches were initially introduced
using solely optical data \citep[e.g.,][]{Gladders2000}, and have
identified hundreds of clusters out to $z \sim 1$. More recently,
the {\it Spitzer} Adaptation of the Red-sequence Cluster Survey
\citep[SpARCS;][]{Wilson2009} has pushed this method to redder
wavelengths and thus to higher redshifts. SpARCS has identified
clusters out to $z\sim1.3$. However, a weakness of this approach
is that it requires the presence of a well-formed red sequence of
early-type galaxies. Such a sequence might not yet exist as we
approach the epoch of cluster formation.

Other field studies such as the IRAC Shallow Survey
\citep[ISCS;][]{Stanford2005, Brodwin2006,Eisenhardt2008} also
used the Infrared Array Camera \citep[IRAC;][]{Fazio2004A} onboard
{\it Spitzer} to expand the sample of galaxy clusters known at
$z>1.2$. In particular, the ISCS identifies clusters on the basis
of photometric redshifts and does not require a red sequence. IRAC
is an extremely sensitive tool for finding massive galaxies at high
redshift since their $4.5\mu$m flux densities remain nearly constant
at $0.7<z<2.5$ due to a negative and favorable $k$-correction
\citep[e.g.,][]{Eisenhardt2008}.

\citet{Papovich2008} used a simple IRAC color criterion (see Section 4) 
to highlight overdensities of high-redshift galaxies in the {\it Spitzer} Wide-Infrared 
Extragalactic \citep[SWIRE;][]{Lonsdale2003} survey and, in doing so, isolated one 
of the highest redshift clusters known to date, ClG~J0218-0510 at $z=1.62$ in the 
{\it XMM}-LSS field of SWIRE. Spectroscopic follow-up confirmed $9$ members \citep{Papovich2010}. 
This same cluster was independently discovered by \citet{Tanaka2010} using photometric 
redshifts. This second team found two concentrations of galaxies at $z_{\rm ph}\sim1.6$, one of them 
associated with extended X-ray emission. They spectroscopically confirmed six members.

These various IRAC surveys have the strong advantage of providing
uniformly selected galaxy cluster samples over wide areas. Such uniform cluster
samples are beneficial for a range of studies, including probing
the formation epoch of the early-type galaxy populations 
\citep[e.g.,][]{Mancone2010, Rettura2011} and statistical
probes of cosmological parameters \citep[e.g.,][]{Vikhlinin2009, 
Stern2010}.

However, clusters are rare objects and finding larger samples of massive high-redshift galaxy 
clusters would require field surveys even wider than the several tens of square-degrees which 
is the current state-of-the-art. Work has been done using targeted cluster searches, i.e.,~focusing 
on regions of the sky suspected to host overdensities of galaxies. For example, powerful, 
high-redshift radio galaxies\footnote[1]{Following Seymour et al.~(2007), we define a HzRG as 
a radio galaxy above a redshift of one with a {\it restframe} $3$~GHz luminosity greater than 
$10^{26}$ W Hz$^{-1}$.} (HzRGs hereafter) are among the most massive galaxies known up to 
very high redshift \citep{Seymour2007} and, as such, are suspected to lie preferentially in overdense 
regions. Studies of HzRG environments have revealed excesses of extremely red objects 
\citep[EROs;][]{Stern2003, Best2003}, line emitters detected through narrow-band imaging 
\citep[][and references therein]{Pentericci2000, Venemans2007} and submillimetre galaxies 
\citep{Stevens2003, DeBreuck2004, Greve2007} in the fields of HzRGs out to $z\sim5$. A 
range of complementary studies have made use of optical through mid-infrared observations 
to isolate candidate cluster members associated with radio galaxies (Kodama et al.~2007; 
Galametz et al.~2009, 2010a; Hatch et al.~2010; Mayo et al.~2012 in press)
\nocite{Kodama2007, Galametz2009B, Galametz2010B, Hatch2011}.

Recently, \citet{Falder2010} extended this cluster selection technique to a large sample of 
high-redshift active galactic nuclei (AGN) at $z\sim1$. Their sample, observed at $3.6\mu$m with 
IRAC, included both radio-loud and radio-quiet AGN. They found excesses of $3.6\mu$m sources
within $300$~kpc of the AGN, as well as evidence for a positive correlation 
between source density and radio power.

In this paper, we use mid-infrared $3.6$ and $4.5\mu$m observations of radio galaxies at 
$1.2<z<3$ to study their environments and identify their possible association with high-redshift 
galaxy clusters. Note that we use the designation `cluster' to refer to both cluster- and 
protocluster-like structures. The paper is organized as follows. The HzRG sample and 
catalog extraction are presented in \S2 and \S3. In \S4, we introduce our IRAC selection 
criterion for high-redshift cluster member candidates. In \S5, we present the counts-in-cell 
analysis used to isolate overdensities associated with HzRGs. We also present the spatial 
distribution of IRAC-selected sources in our most promising high-redshift cluster candidates 
in \S6. We discuss our $z>3$ fields in \S7 and present our conclusions in \S8. Throughout, 
we assume a $\Lambda$CDM cosmology with $H_0 = 70$ km s$^{-1}$ Mpc$^{-1}$, 
$\Omega_m = 0.3$ and $\Omega_{\Lambda} = 0.7$. Magnitudes and colors are expressed 
in the AB photometric system unless stated otherwise.

\section{Radio galaxy sample and IRAC data}

Our primary HzRG sample comes from the {\it Spitzer} High-Redshift Radio Galaxy survey (SHzRG; 
Seymour et al.~2007, De Breuck et al.~2010). Most of the sources were observed with IRAC during 
Cycle 1 (PID 3329; PI Stern) with typical integration times of $120$s in all four IRAC bands (see 
Table~\ref{targets}). Some fields were imaged more deeply using guaranteed time observations 
(see Table~\ref{targets2}). Details on the IRAC reductions can be found in \citet{Seymour2007} and 
\citet{DeBreuck2010}. Mayo et al.~2012 (in press) report on the MIPS $24\mu$m environments of $64$ 
of these fields.

The initial SHzRG sample was designed to homogenously cover the $L_{\rm 3 GHz}$ radio 
luminosity - redshift plane. More recent work on the SHzRG sample, including this paper, instead 
use $L_{\rm 500 MHz}$, which is a more isotropic measure of AGN power \citep{DeBreuck2010}. 
This leads to a slightly less uniform distribution in the $L_{\rm 500 MHz}$ radio luminosity 
- redshift plane (see Fig.~2 of De Breuck et al.~2010). We will keep this in mind by using appropriate
statistical tests later in the paper (see  Section 5.3).


In 2009, we obtained deeper $1600$s IRAC observations of a sample of $10$ radio galaxies (see 
Table~\ref{targets2}, PID 60112; PI Hatch) at $3.6$ and $4.5\mu$m. Six of these fields were already 
part of the initial shallower sample.
In this paper, we make use of the deeper data when available. These deeper data were reduced as in 
\citet{Seymour2007}. 

The goal of our study is to isolate galaxy clusters at high redshift using only mid-infrared data. 
As described below, the IRAC criterion applied in this work (see Section 4) is optimal for 
isolating galaxy structures at $1.2<z<3$. We therefore concentrate our analysis on the $48$ 
HzRG fields in this redshift range. We also analyze the $25$ HzRG fields at $z < 1.2$ and 
$3 < z < 5.2$ from our sample in an identical manner and use these fields as a control sample.

\section{Catalog extraction and flux limits}

In order to study the mid-infrared environments of our HzRG sample, we analyzed the available
{\it Spitzer} data using standard techniques.

We restricted the analysis to areas covered for at least $60$s in both IRAC channel 1 ($3.6\mu$m) 
and 2 ($4.5\mu$m). The central part ($\sim1\arcmin \times 1\arcmin$) exposure time is $120$s. We 
checked the images by eye and flagged zones affected by artifacts (e.g.,~scattered light). 
Source extraction was done using SExtractor \citep{Bertin1996} in dual image mode, using the 
$4.5\mu$m frame as the detection image. We used SExtractor parameters from \citet{Lacy2005} 
which have been optimized for analysis of IRAC data, and measured photometry in 3\arcsec\ diameter 
apertures. These magnitudes were then corrected to total magnitudes using corrections determined by 
the IRAC instrument team (M. Lacy, private communication). Specifically, the aperture corrections applied 
were $1.68$ and $1.81$ for channels 1 and 2 respectively.

Limiting flux densities for each image were determined from randomly placed $3\arcsec$ diameter apertures.
For the SHzRG fields, the limiting flux densities were determined on the area covered by $120$s exposure 
which extends beyond a radius of $1\arcmin$ from the HzRG i.e.,~the size of the cell on which this work
is conducted. To allow for uniform analysis of our large data set, we adopt a conservative flux density cut 
corresponding to the $5\sigma$ depth of our shallowest data, i.e.,~$f_{3.6}=11.0\mu$Jy ($[3.6]=21.3$) 
and $f_{4.5}=13.4\mu$Jy ($[4.5]=21.1$)\footnote[2]{Magnitudes in IRAC channels 1 and 2 are 
indicated by $[3.6]$ and $[4.5]$, respectively, and $m_{\rm AB}=23.9-2.5$ log($f_{\nu}/1 \mu$Jy).}.

Recently, \citet{Mancone2010} derived the $3.6$ 
and $4.5\mu$m luminosity functions of galaxy clusters out to $z\sim1.5$. They found $[3.6]^*\sim20.0$ 
 and $[4.5]^*\sim20.1$ for clusters at $z\sim1.2$. Similarly, \citet{Strazzullo2006} derived the
$K_s$-band luminosity function of clusters at $z\sim1.2$. Using the \citet{Bruzual2003} model of a 
$2$~Gyr-old single burst galaxy with an exponentially declining star formation history with $\tau=0.1$, we 
find that $K-[3.6]\sim0.3$ at $z\sim1.2$ with a nearly constant $K-[3.6]$ color at $1<z<2$. \citet{Strazzullo2006} 
find that $K^*_s\sim20.5$ at $z\sim1.2$, corresponding to $[3.6]^*\sim20.2$, in agreement with
the work of \citet{Mancone2010}. To our adopted limiting magnitude of $[3.6]=21.3$, our shallow $120$s
observations are therefore sensitive to galaxies one magnitude fainter than $L^*$ at $z\sim1.2$. As shown in
\citet{Mancone2010}, $m^*$ should remain relatively constant out to $z\sim2$ assuming 
cluster galaxies form at $z\geq2.5$.

\section{Color selection of cluster member candidates}

\begin{figure}
\begin{center}
\includegraphics[width=9cm,bb=50 50 500 500]{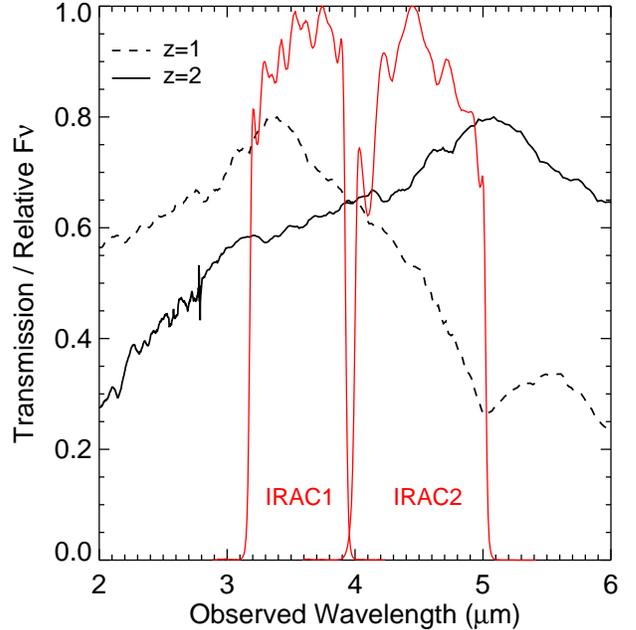}
\end{center}
\caption{Spectral energy distribution of a 2~Gyr-old galaxy redshifted to $z=1$ (dashed line) 
and $z=2$ (solid line) assuming a rapid ($\tau=0.1$~Gyr) exponentially declining star-formation
history. The strongest feature in this wavelength range is the {\it restframe} $1.6\mu$m stellar bump.
Transmission curves for IRAC channel 1 and 2 are overplotted for reference.}
\label{temp}
\end{figure}

One of the main spectral features in the spectral energy distributions
(SEDs) of galaxies is the $1.6\mu$m bump, caused by a minimum in
the opacity of the H$^-$ ion which is present in the atmospheres
of cool stars \citep{John1988}. This bump, seen in the SEDs of all
normal galaxies, has often been used as an efficient photometric
redshift indicator \citep[e.g.,~][and references therein]{Simpson1999,
Sorba2010}. Fig.~\ref{temp} shows the position of the $1.6\mu$m
bump for galaxies at $z=1$ and $z=2$ relative to the IRAC $3.6$ and
$4.5\mu$m bands. The bump enters the IRAC bands at $z\sim1$ and
shifts beyond $3.6\mu$m at $z\sim1.2$. This causes lower redshift
galaxies to have blue $[3.6]-[4.5]$ colors, while galaxies above
$z\sim1.2$ become red across these passbands. 

Combining a wide variety of composite stellar population models and
spectroscopy of high-redshift galaxies in the GOODS-S and AEGIS
DEEP2 fields, \citet{Papovich2008} confirmed that $[3.6]-[4.5]>-0.1$
(i.e.,~$f_{3.6}/f_{4.5}<1.1$) is very efficient at isolating $z>1.2$
galaxies. They showed that only a minority population of $0.2<z<0.5$
strongly star-forming galaxies with a strong warm dust contribution
have such red colors and that about $80$\% of galaxies with
$[3.6]-[4.5]>-0.1$ are at $z>1$. Specifically, based on redshifts
from the DEEP2 redshift survey, \citet{Papovich2008} show that $\sim
50$\% of galaxies at $z > 1.1$ have red IRAC colors, with this
percentage increasing to $\sim 75$\% for $z > 1.2$ galaxies and
reaching $\sim 90$\% for $z > 1.3$ galaxies. In the later analysis
we use this color cut to study the fields of HzRGs at $z > 1.2$,
though we caution the reader than the efficiency of the selection
criterion is diminished in the $1.2 < z < 1.3$ range. At any rate,
only one HzRG from our sample, 3C266 at $z = 1.275$, is in this
redshift range. Finally, we note that, contrary to other
high-redshift galaxy selection techniques such as the $BzK$ selection
for $z>1.4$ galaxies \citep{Daddi2000} or the near-infrared selection
techniques for $z>1.6$ galaxies \citep{Kajisawa2006, Galametz2010a},
this single IRAC color criterion does not permit the segregation
between galaxy types since it is based on a spectro-photometric
property of essentially all galaxy populations \citep{Papovich2008}.


This red IRAC color criterion will identify a few additional astronomical populations, but none are 
expected to be significant contaminations. Most stars have $[3.6]-[4.5]\sim-0.5$ (i.e.,~Vega color of zero). 
\citet{Stern2007} shows that only brown dwarfs cooler than spectral type T3 have such red colors 
($[3.6]-[4.5]>-0.07$) and such objects will be rare in a flux-limited survey at our depth. Extremely 
dusty stars, such as asymptotic giant branch (AGB) stars, will also have red IRAC colors 
\citep[e.g.,][]{Eisenhardt2010}. However, they are likewise expected to be quite rare in a 
flux-limited survey primarily targeted at high Galactic latitudes.

At all redshifts, most powerful AGN ($>95$\%) have $[3.6]-[4.5]>-0.1$ and 
will be selected by this IRAC criterion. \citet{Stern2005} measure a surface 
density of $275$ AGN per deg$^2$ in the Bo\"{o}tes field down to flux density 
limits of $f_{3.6}=12.3\mu$Jy and $f_{4.5}=15.4\mu$Jy. Using IRAC catalogues 
available for the Bo\"{otes} field (see Section 5.2 for details), we derive a density of 
$\sim28,700$ red IRAC-selected sources per deg$^2$ to the same depth. We 
therefore expect contamination by AGN to represent less than $1$\% of sources 
with $[3.6]-[4.5]>-0.1$ at the depth of our survey.

\begin{figure}
\begin{center}
\includegraphics[width=9cm,bb=0 0 500 500]{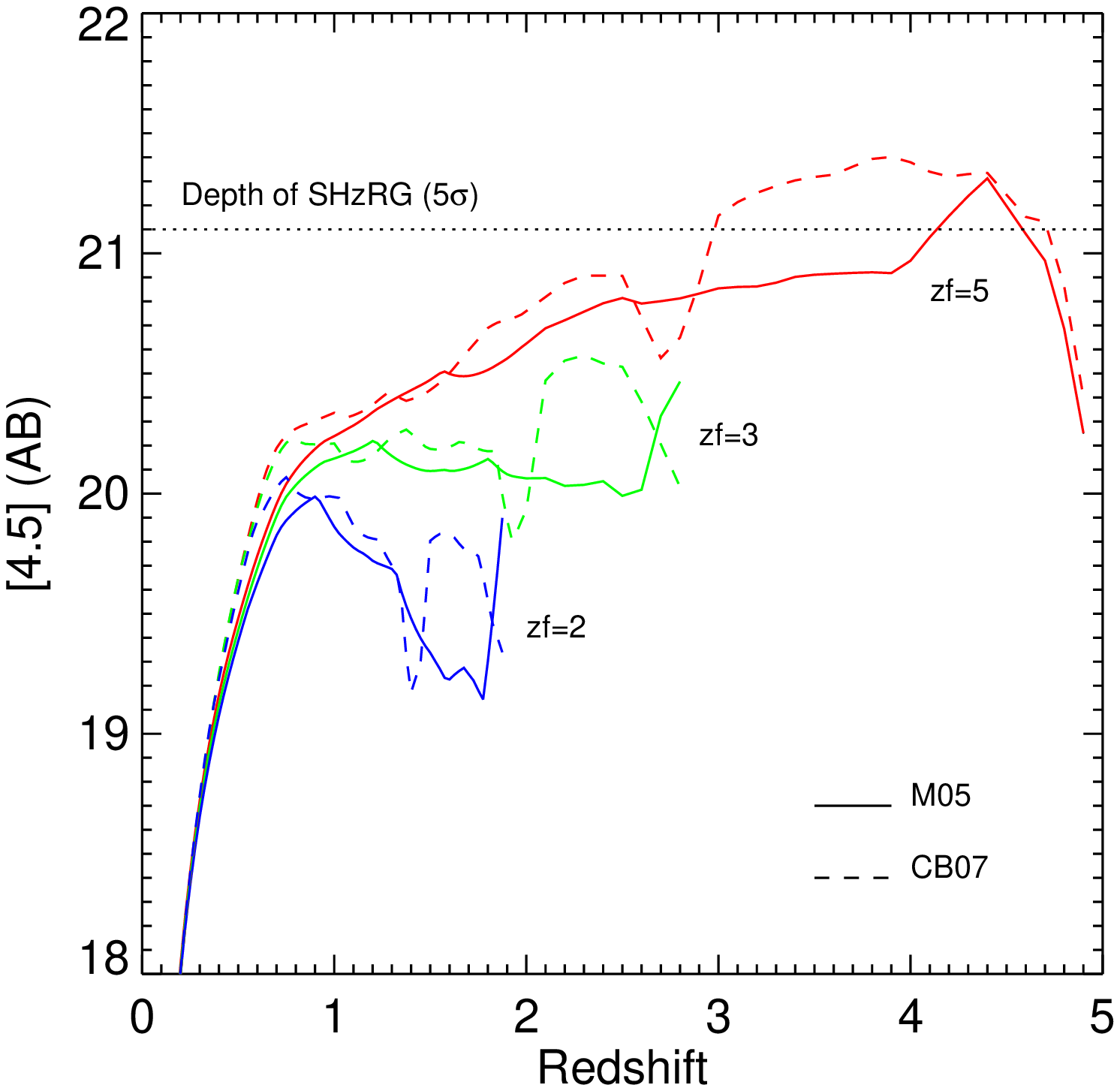}\\
\includegraphics[width=9cm,bb=0 0 500 500]{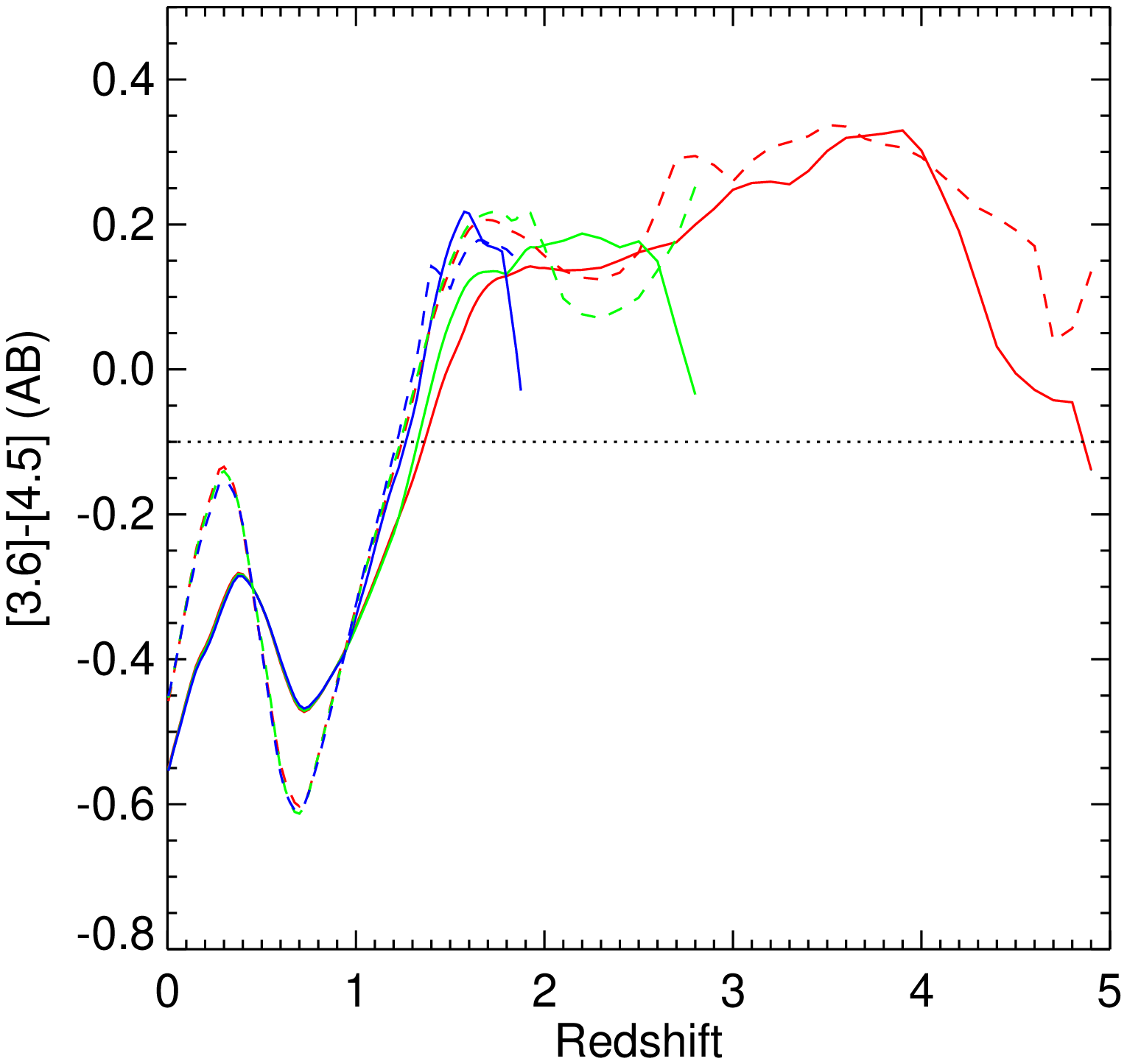}
\end{center}
\caption{Evolution of $4.5\mu$m flux density (top panel) and
$[3.6]-[4.5]$ color (bottom) versus redshift for single stellar
populations with a range of formation redshift, $z_f=2$ (blue), $3$
(green) and $5$ (red). Rapid ($\tau=0.1$~Gyr) exponentially declining
star-formation history models, from Maraston et al.~2005 (M05; solid
line) and Charlot \& Bruzual 2007 (CB07; dashed), were generated
with EZ Gal (assuming $[4.5]_{\rm Vega}=16.75$ at $z\sim0.7$; see
text for details). At $z>1$, $[4.5]$ is relatively independent of
redshift out to the formation epoch. The dotted lines indicate the
$[4.5]$ depth (top) and the $[3.6]-[4.5]>-0.1$ color criterion (bottom)
adopted in this work.}
\label{conor}
\end{figure}

According to models for passively evolving stellar populations
formed at high-redshift, negative $k$-corrections provide a nearly
constant $4.5\mu$m flux density at $z>0.7$ and up to high redshifts.
Fig.~\ref{conor} shows the evolution of the $4.5\mu$m flux
density and the $[3.6]-[4.5]$ color as a function of redshift for
various formation redshifts. We make use of the publicly available
model calculator EZ Gal\footnote[3]{\tt www.mancone.net/ezgal/model}.
We use the \citet{Maraston2005} and the Charlot \& Bruzual~2007 ---
an update of \cite{Bruzual2003} --- models for single stellar
populations assuming solar metallicity and a Salpeter initial mass
function. The $4.5\mu$m flux density is normalized to match the
observed $m^*_{4.5}$ of galaxy clusters at $z\sim0.7$, $[4.5]_{\rm
Vega}=16.75$. \citet{Mancone2010} studied a large sample of clusters
up to $z\sim1.5$ to derive the $4.5\mu$m luminosity function and
found that the luminosity function is consistent with a formation
redshift $z_f\sim2-3$. Fig.~\ref{conor} shows that for $z_f=3$, the
$4.5\mu$m flux density is relatively constant at $z>1$. It even
decreases (brightens) with redshift as one approaches
the galaxy formation epoch. 

\citet{Mancone2010} models could not simultaneously match the low
and high-redshift data; they found that $m^*$ becomes fainter at
the highest redshifts probed, interpreted as a possible evidence
for galaxy assembly at $z\ge1.5$. Note, however, that the
\citet{Mancone2010} uncertainties increase with redshift, with few
of the highest redshift clusters spectroscopically confirmed, and
larger systematic uncertainties in subtracting the foreground/background
galaxy populations. Numerous authors have investigated the epoch
of early-type galaxy assembly, and the results span a wide range
of formation redshifts \citep[e.g.,][]{VanDokkum2008, Eisenhardt2010}.
A larger sample of high-redshift galaxy clusters is clearly needed
to further investigate this question. 

In the following, we assume a uniform $[3.6]-[4.5]>-0.1$ color
criterion to select high-redshift candidates in the fields of 48
radio galaxies at $1.2 < z < 3$. We consider sources detected to
the $5\sigma$ depth of our SHzRG sample ($[4.5]=21.1$). We include
sources with $[3.6]-[4.5]>-0.1$ with a $3.6\mu$m magnitude fainter
than the limits of our SHzRG survey. For such sources, we uniformly
assign a lower limit of $[3.6]=21.3$. Sources selected by these
criteria are referred to as `red IRAC-selected sources' or simply
as `IRAC-selected sources' in the rest of the paper. 

As seen in Fig.~\ref{conor}, our selection criteria should robustly
identify clusters across this whole redshift range, assuming cluster
galaxy formation and assembly is at similar or higher redshift.
The aim of the current work is to test this hypothesis. Current
literature only studies evolved populations in clusters out to $z
\sim 1.3$, with less certain results for cluster candidates at
higher redshifts. Extensive literature from multiple observing
programs infer the cluster formation epoch is at higher redshift,
with results ranging over the $2 \simlt z_f \simlt 5$ range. Given
the current state of knowledge, the high redshift end of our analysis
is somewhat arbitrary; we could have been more conservative and
chosen a lower redshift cut-off, or, in principle, we could have
considered a higher redshift cut-off. If we do not find rich fields
around our highest redshift HzRGs, multiple explanations are possible.
It could mean that clusters do not exist at those redshifts, it
could be due to small number statistics, or it could imply issues
with the simple evolutionary model we are testing in Fig.~\ref{conor},
a model that is remarkably effective at describing cluster data at
$z < 1.3$. In particular, galaxy mergers might cause an evolution
of $m^*$ with redshift such that we become less sensitive to clusters
at the high-redshift end of our sample. Such a result would be of
interest.

\section{Counts-in-cell analysis}

\subsection{Counts-in-cell in SWIRE}

\begin{figure*}
\begin{center} 
\includegraphics[width=14cm]{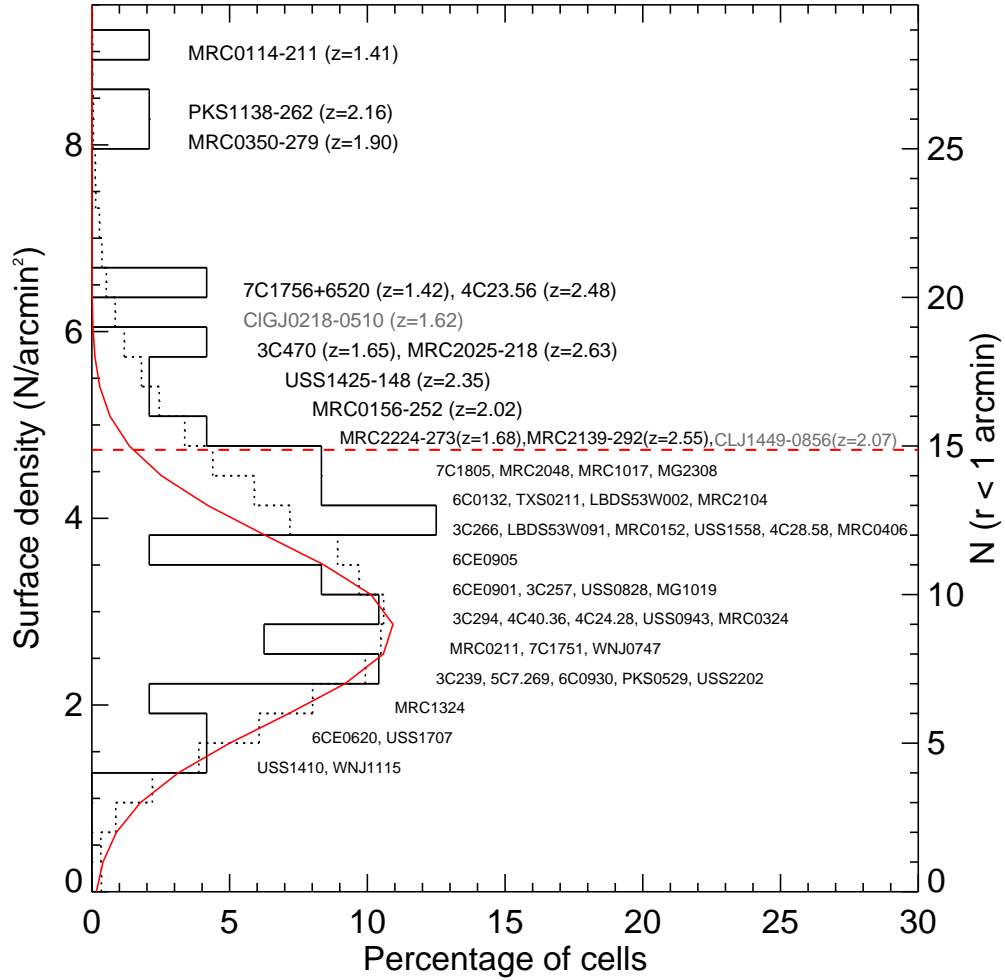}
\end{center}
\caption{Distribution of the surface density of $[3.6]-[4.5]>-0.1$ sources in $20,000$ cells 
of $1\arcmin$ radius randomly generated in the six SWIRE fields to the flux density limits of our sample
(dotted histogram). The corresponding number of sources per cell is indicated on the right axis. A Gaussian fit of the 
lower part of the distribution is shown by the red curve and the $2\sigma$ deviation above the mean of 
this distribution by the horizontal dotted line. Each radio galaxy is placed at the vertical axis 
location corresponding to the surface density of IRAC-selected sources found within $1\arcmin$ 
from the radio galaxy. The solid histogram shows the percentage of radio galaxy fields for a given density. 
Two examples of known high-redshift galaxy clusters (CLG~J0218-0510 and Cl~J1449+0856)
are also indicated by the gray labels (see \S5.2).}
\label{cell}
\end{figure*}

We use a counts-in-cell analysis to identify overdensities of IRAC
red galaxies associated with HzRGs. As a reference field, we analyze
the SWIRE 
\citep[SWIRE;][]{Lonsdale2003} survey. With its typical $120$s
exposure, the SWIRE survey reaches slightly deeper than our HzRG
data, primarily because it exclusively targets low-background regions
of the sky. SWIRE covers $\sim50$ deg$^2$, split into six independent
fields which mitigates the effects of cosmic variance.

To ensure consistency with our analysis of the radio galaxy fields,
we build our own catalogues for all six SWIRE fields. As described
in Section 3, we use the $4.5\mu$m image for source detection and derive
photometry from $3\arcsec$ diameter apertures, corrected to total
magnitudes following \citet{Lacy2005}.

At the depth of our survey, we detect the bright end of the galaxy
luminosity function, which preferentially populates cluster cores.
We therefore adopt a cell size of $1\arcmin$ radius, corresponding
to $\sim0.5$~Mpc at $1<z<3$. This distance matches typical cluster
$r_200$ sizes for $z>1.3$ mid-infrared selected galaxy clusters
\citep[e.g.,][]{Wilson2009}.

We generate $20,000$ independent (i.e.,~non-overlapping) circular
cells of $1\arcmin$ radius, randomly distributed over the six SWIRE
fields. Fig.~\ref{cell} shows the histogram of the surface density
of red IRAC-selected sources per cell. The horizontal axis shows
the percentage of cells corresponding to a specific surface density.
The shape of the distribution is a Gaussian strongly skewed towards
higher density cells. We fit the lower half of the distribution
(i.e.,~the distribution of the lower density regions) by a Gaussian
(iteratively clipping at $2\sigma$). The best fit values give a
Gaussian centered at $\langle N \rangle=2.80$ galaxies per arcmin$^2$
($\sim8.8$ sources per cell) with $\sigma_N=0.97$. We define an
overdensity of IRAC-selected sources when a cell is denser than
$N+2\sigma_N\sim4.74$ galaxies per arcmin$^{2}$ --- i.e.,~ when at
least $15$ sources are found within a cell of $1\arcmin$ radius.
This criterion, for which visual inspection typically registers
interesting spatial segregation, is matched by only $9\%$ of the
cells in SWIRE.

\subsection{Tests on known $z>1$ galaxy clusters}

Before applying our newly defined criterion, we test its efficiency on published
clusters at $z>1$ observed with IRAC. We first study two of the highest redshift 
clusters known to date: ClG~J0218-0510 at $z=1.62$ \citep{Papovich2010, Tanaka2010} 
and Cl~J1449+0856 at $z=2.07$ \citep{Gobat2011}.

ClG~J0218-0510 is located in the XMM field of SWIRE. We select sources with $[3.6]-[4.5]>-0.1$ 
in the field of ClG~J0218-0510 using the catalogues we derived from the SWIRE data (see Section 5). 
Cl~J1449+0856 was discovered in the CNOC~1447+09 field \citep{Yee2000} and was
also first identified as an overdensity of red IRAC-selected galaxies. To date, $11$ cluster members 
have been spectroscopically confirmed \citep{Gobat2011}. We retrieved the IRAC $3.6$ 
and $4.5\mu$m reduced post-BCD images of the CNOC~1447+09 field from the {\it Spitzer} archive (P.I. Fazio) 
and derived our own catalogue in an identical manner as for the HzRG targets.

\begin{figure}
\begin{center}
\includegraphics[width=8cm]{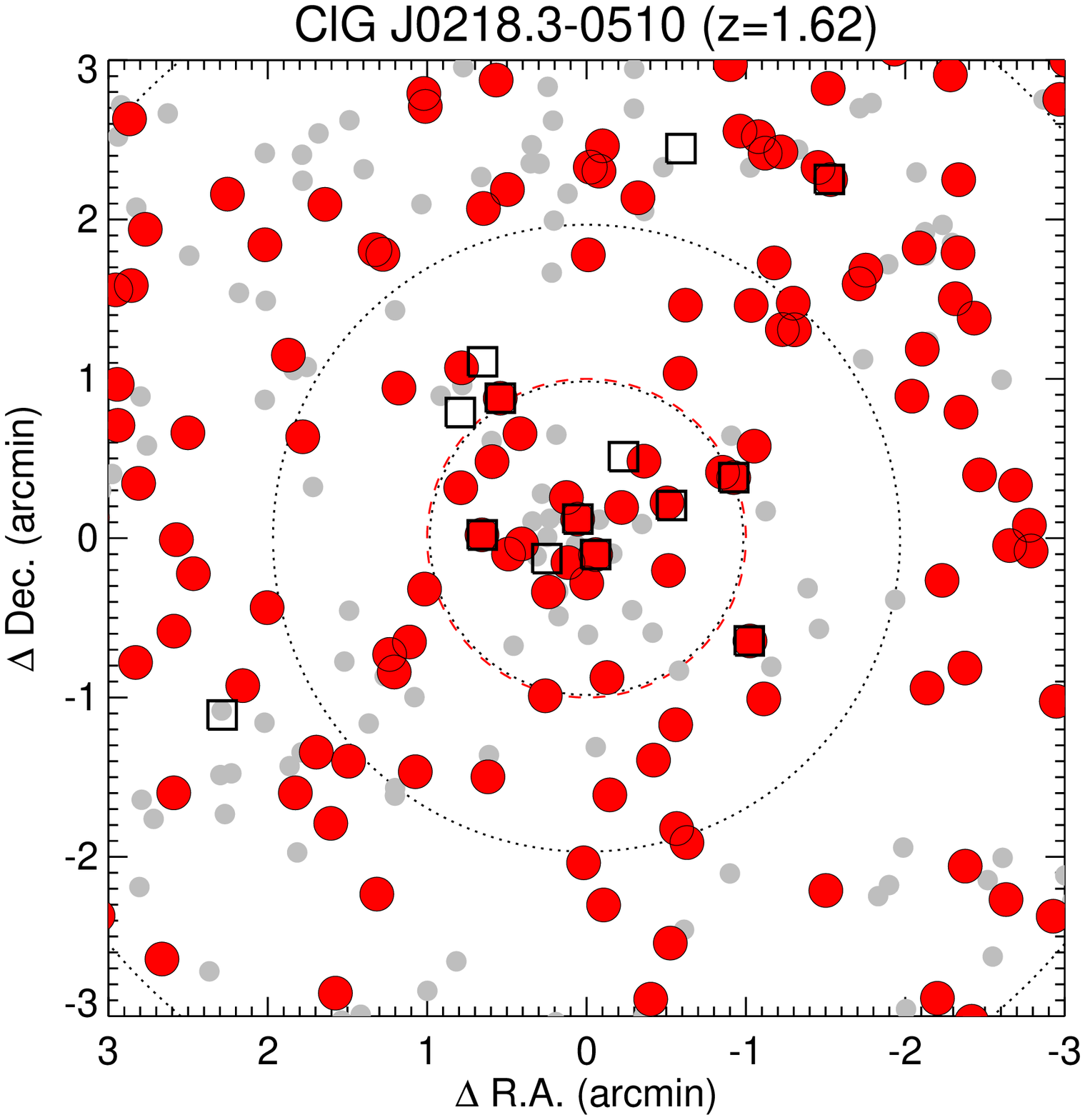}
\includegraphics[width=8cm]{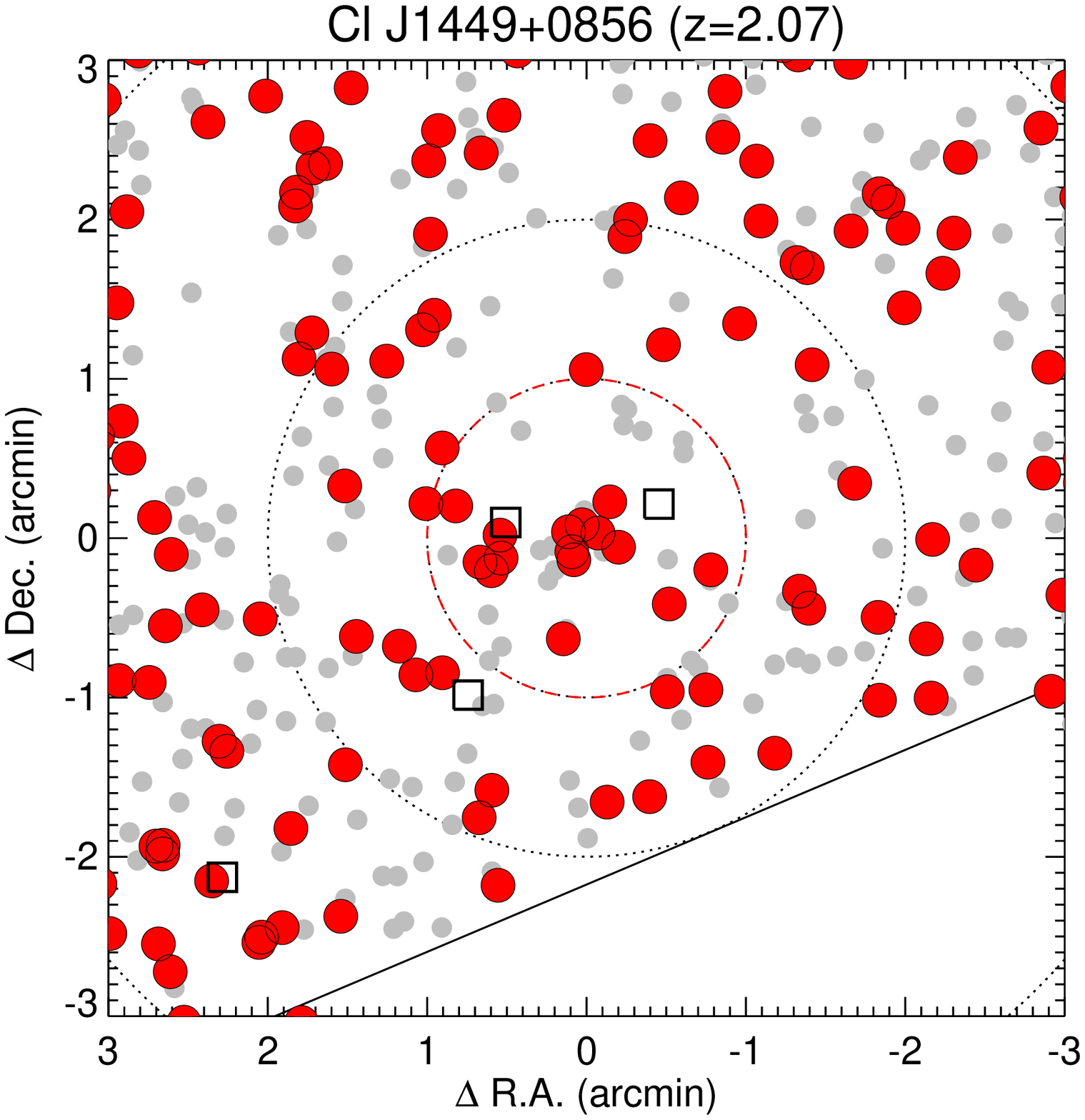}
\end{center}
\caption{Spatial distribution of sources with $[3.6]-[4.5]>-0.1$ in the fields of two
spectroscopically confirmed galaxy clusters: ClG~J0218.3-0510 ($z=1.62$; top) and 
Cl~J1449+0856 ($z=2.07$; bottom; the solid line marks the limit of the area covered by {\it Spitzer}).
Small gray circles indicate sources down to the $5\sigma$ depth of the original IRAC data 
while larger red symbols indicate sources detected to the 
flux limits of our SHzRG data. Dotted circles illustrate $0.5$, $1$ and $2$~Mpc radii at $z=1.62$ 
and $z=2.07$. A $1\arcmin$ radius cell ($\sim0.5$Mpc) centered on the adopted cluster center is shown 
by the red dashed circle. Spectroscopically confirmed members are 
indicated by black squares. North is up; East is to the left.}
\label{ClGJ0218}
\end{figure}

For both cluster fields, we select sources with $[3.6]-[4.5]>-0.1$. Fig.~\ref{ClGJ0218} shows the spatial 
distribution of these red sources in the $6\arcmin \times 6\arcmin$ field around ClG~J0218-0510
(top) and Cl~J1449+0856 (bottom panel). Large red symbols account for sources detected within the limits 
applied to our SHzRG sample and smaller gray dots account for sources within the $5\sigma$ limits of 
SWIRE ($f_{3.6}=3.7\mu$Jy and $f_{4.5}=5.4\mu$Jy) and CNOC~1447+09 ($f_{3.6}=2.3\mu$Jy 
and $f_{4.5}=3.8\mu$Jy). We also show the spectroscopically confirmed cluster members (black squares)
for ClG~J0218-0510 \citep{Papovich2010, Tanaka2010} and Cl~J1449+0856 (R.~Gobat, private communication). 
All of the confirmed members detected at $3.6$ and $4.5\mu$m have $[3.6]-[4.5]<-0.1$ though some are below
the $5\sigma$ limits plotted in Fig.~\ref{ClGJ0218}. 

Nineteen red IRAC-selected galaxies are found within $1\arcmin$ of the assigned center of ClG~J0218-0510 
and $15$ for Cl~J1449+0856. Both clusters are clearly seen as compact concentrations of red galaxies
and are selected as overdense fields by our IRAC criteria (see also Fig.~\ref{cell}, gray labels).

An overdensity of EROs was also recently detected in the field of the quasar 3C270.1 at $z=1.53$ 
\citep{Haas2009} suggesting that the quasar is associated with a high-redshift galaxy structure.
We retrieved the IRAC post-BCD $3.6$ and $4.5\mu$m images of this field from the {\it Spitzer} archive 
and extracted the corresponding $4.5\mu$m selected catalogue. Twenty-four red IRAC-selected sources are 
found within $1\arcmin$ of the quasar. This strengthens past results that the quasar is 
part of a high-redshift cluster.

Recently, \citet{Eisenhardt2008} identified a large sample of $335$ mid-infrared selected 
galaxy clusters in the Bo\"{o}tes field, including five clusters spectroscopically confirmed 
at $1.2 < z < 1.5$. The Bo\"{o}tes field was covered by the {\it Spitzer} Deep, Wide-Field Survey 
\citep[SDWFS;][]{Ashby2009}, an IRAC survey of $8.5$ deg$^2$ with $5\sigma$ depths of $22.9$ 
and $22.4$ in $3.6$ and $4.5\mu$m, respectively. We make use of the publicly available 
$4.5\mu$m-selected catalogue of the SDWFS Data Release 1 to derive the number 
of IRAC-selected sources within 
$1\arcmin$ of the cluster centers. All five $z>1.2$ confirmed galaxy clusters have $15$ or more IRAC-selected 
sources in the studied cell and would be selected as overdense fields by our IRAC criteria. 
Of the $40$ spectroscopically confirmed members of these clusters at $z>1.2$ 
detected at $4.5\mu$m (Stanford et al.~2005, Brodwin et al.~2006, 
Eisenhardt et al.~2008)\nocite{Brodwin2006,Stanford2005}, $25$ ($63$\%) have $[3.6]-[4.5] > -0.1$. 
Considering only clusters members at $z > 1.3$, all $21$ galaxies have colors 
consistent with the red IRAC criterion within the photometric errors.
Considering the $z < 1.2$ clusters reported in Eisenhardt et al. (2008), only the cluster at $z =
1.161$ would be selected as an overdense field by our red IRAC selection criterion. These results 
confirm that our IRAC criterion begins to become effective at identifying galaxy clusters at $z > 1.2$ and 
is quite robust at identifying rich structures at $z > 1.3$.

A large number of high-redshift galaxy cluster candidates were also recently found in SWIRE
by the SpARCS survey. We further test our selection criterion on spectroscopically confirmed 
SpARCS clusters and recover their two highest redshift galaxy clusters (to date): 
SpARCS J003550-431224 \citep[$z = 1.34$;][]{Wilson2009} --- a very dense structure with 
$35$ IRAC-selected sources found within $1\arcmin$ of the assigned cluster center --- and 
SpARCS J161037+552417 \citep[$z = 1.21$;][]{Demarco2010} with $19$ red IRAC-selected 
sources. We note that the two $z<1.2$ clusters published in Demarco et al.~2010 (at $z = 0.871$ 
and $z = 1.161$) were not recovered.


\subsection{Radio galaxy fields}

We now determine the number of red IRAC-selected sources within $1\arcmin$ of each radio galaxy. 
Each field is shown in Fig.~\ref{cell} at the position corresponding to the number of IRAC-selected 
sources found within $1\arcmin$ of the radio galaxies. The density for each field is also reported 
in Tables 2 and 3 (last column). The solid histogram shows the percentage of radio galaxy fields
corresponding to a given density. We find that radio galaxies preferentially lie in medium 
to dense regions, with $73\%$ ($\pm12$\%) of the targeted fields denser than the mean of the Gaussian fit to 
the SWIRE fields. Eleven fields i.e.~$23$\% ($\pm7$\%) are found with more than $15$ red 
IRAC-selected sources within $1\arcmin$ of the central radio galaxy. We conduct a two-sided
Kolmogorov-Smirnov statistic between the distributions of densities of IRAC-selected sources
in SWIRE and in the HzRG cells and find a probability of only $1.3$\% that the two distributions
are drawn from the same distribution.

From the $20,000$ cells we produced in SWIRE, we generated $100,000$ random samples
of $48$ cells and derived the fraction of these sub-samples that have at least $11$ $2\sigma$-overdense 
cells. We find that only $0.3$\% of the sub-samples have a fraction of overdense cells 
comparable to our radio galaxy sample, confirming that radio galaxies lie preferentially 
in dense regions.

We discuss our $11$ overdense fields in more detail in Section 6. At our bright magnitude limits,
we expect to isolate high-redshift galaxy structures that contain many bright 
($M^*+1$ or brighter) galaxies in their cores. We do not rule out the possibility that the 
other radio galaxies may be associated with less massive and/or less compact galaxy structures.

\begin{figure}
\begin{center} 
\includegraphics[width=8.5cm]{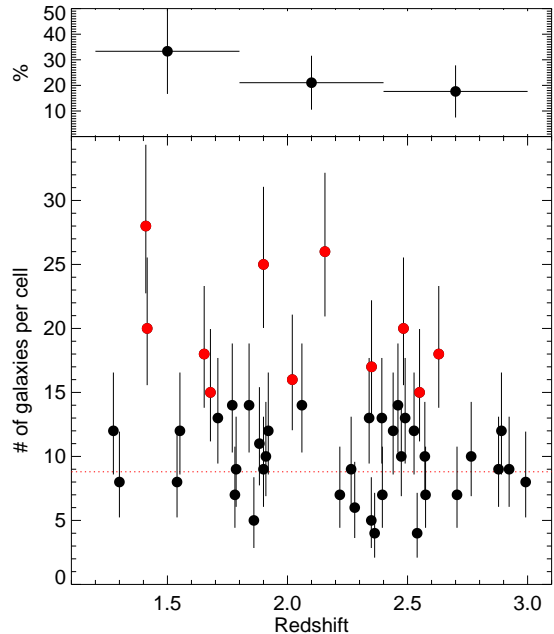}
\end{center}
\caption{Number of $[3.6]-[4.5]>-0.1$ sources within $1\arcmin$ of the radio galaxies 
(and within the $5\sigma$ $4.5\mu$m limit of our sample) as a function of redshift. 
Overdense environments are marked in red. The horizontal red line shows the median value 
($8.8$) for a $1\arcmin$ radius cell in SWIRE. The top panel shows the percentage of 
overdense fields relative to the total number of fields per bin of redshift.} 
\label{zlum1}
\end{figure}

\begin{figure}
\begin{center} 
\includegraphics[width=8.5cm]{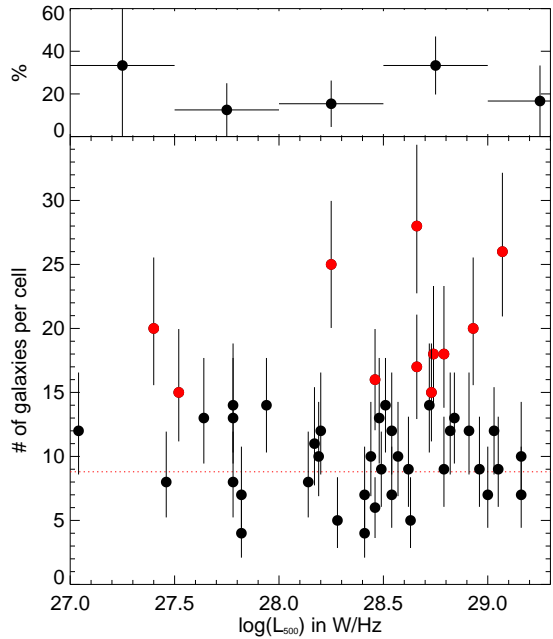}
\end{center}
\caption{Number of $[3.6]-[4.5]>-0.1$ sources within $1\arcmin$ of the radio galaxies 
as a function of radio luminosity. Symbols are as in Fig.~\ref{zlum1}. 
We note that the majority of our cluster candidates are found around HzRGs with 
$L_{\rm 500MHz}>10^{28.6}$~W Hz$^{-1}$ suggesting that more powerful radio 
galaxies reside in denser environments.}
\label{zlum2}
\end{figure}

The sample of radio galaxies covers a wide range of redshift as well as a wide range of radio 
luminosity. We next consider a possible dependance of overdensity with these two parameters. 
Fig.~\ref{zlum1} and Fig.~\ref{zlum2} show the counts-in-cell around 
the radio galaxies versus redshift and radio luminosity, respectively. We adopt Poissonian 
errors for the counts-in-cell and use the \citet{Gehrels1986} small number approximation
for Poisson distribution. 
The inset on top of each figure shows the percentage of overdense fields compared to the 
total number of fields per bin of redshift or radio luminosity.


No significant correlation is observed between overdensity and redshift (Fig.~\ref{zlum1}).
As far as a correlation with radio power is concerned (Fig.~\ref{zlum2}), we note that $35$\% ($\pm13$\%)
of radio galaxies with $L_{\rm 500MHz}>10^{28.6}$~W Hz$^{-1}$ are found in overdense
fields, compared to $23$\% of the full sample. This could indicate a positive correlation 
between overdensity and radio power, a result suggested by past studies \citep{Miley2008, Falder2010}.
However, the Spearman rank correlation coefficient between overdensity and radio power 
is as low as $0.071$. As mentioned in Section 2, there may be a small degeneracy between 
$L_{\rm 500 MHz}$ and redshift in our sample. A study of the Spearman partial rank correlation 
coefficient --- which takes the correlation of two variables with a third related variable 
\citep{Macklin1982} --- still do not show any significant correlation between overdensity and radio 
luminosity. This result is consistent with the absence of correlation observed between overdensity 
and the density of $24\mu$m MIPS-selected sources in the same HzRG sample (Mayo et al.~2012 
in press). We conclude that a larger sample, reaching radio luminosities of several orders of 
magnitude lower, is required to study (and quantify) a possible correlation.

\subsection{Analysis of HzRG fields with deeper IRAC data}

We also examine fields with at least $1600$s IRAC exposures. We adopt 
correspondingly deeper cuts for the deeper sample ($f_{4.5}=3.4\mu$Jy; $[4.5]=22.6$). 
Similar to the shallower analysis, we select sources using the criterion $[3.6]-[4.5]>-0.1$.
For sources below the deeper $3.6\mu$m flux density limits, we uniformly assign 
a lower limit $[3.6] = 23.0$. The adopted depth corresponds to $\sim M^*+3$. 
We derive densities of red IRAC-selected sources within $1\arcmin$ of the radio galaxies to 
these deeper flux limits (Table~\ref{dens}). We find results consistent with our 
analysis at shallower cuts, recovering PKS1138-262 ($z=2.156$), MRC0156-252 ($z=2.016$) 
and MRC0350-279 ($z=1.900$) as the densest fields; compared to the average surface 
density of red IRAC-selected sources in the other radio galaxy fields, the environments of 
these three radio galaxies are at least 1.5 times richer.

\begin{table}[!t]
\caption{Surface density of IRAC-selected sources for the deeper fields.}
\label{dens}
\centering
\begin{tabular}{l | c }
\hline
HzRG	&	Density		\\
		&	(arcmin$^{-2}$)	\\
\hline
\hline
PKS1138-262 ($z=2.156$)	&	$18.1 \pm 2.4$	\\	
MRC0156-252 ($z=2.016$)	&	$17.8 \pm 2.4$	\\	
MRC0350-279 ($z=1.900$)	&	$17.5 \pm 2.4$	\\	
MRC1017-220 ($z=1.768$)	&	$14.3 \pm 2.1$	\\	
MRC2104-242	($z=2.491$) 	&	$14.0 \pm 2.1$ \\	
MRC2139-292	($z=2.550$) 		&	$13.7 \pm 2.1$	\\		
MG2308+0336 ($z=2.457$)	&	$12.1 \pm 2.0$	\\
USS1425-148	($z=2.349$) 	&	$12.1 \pm 2.0$	\\
MRC0324-228	($z=1.894$) 	&	$11.5 \pm 1.9$	\\
LBDS53W002 ($z=2.393$) 	&	$10.8 \pm 1.9$	\\
MRC0406-244	($z=2.427$) 	&	$10.8 \pm 1.9$	\\
MRC1324-262	($z=2.280$) 		&	$8.6 \pm 1.7$	\\
\hline
\end{tabular}
\end{table}

\section{Confirmed clusters and new cluster candidates}

Eleven radio galaxy fields present significant overdensities of red sources.
Fig.~\ref{radec} shows the spatial distribution of the IRAC-selected sources 
for these $11$ cluster candidates. Small gray and large red dots indicate
red IRAC-selected sources detected within the $3\sigma$ limits of each individual field and to the depth
of our survey, respectively. These fields present a very diverse spatial distribution of red sources. 
Some fields show clumps (e.g.,~USS1425-148, MRC0350-279) while others show more 
filamentary structures (e.g.,~MRC0114-211, 3C470, PKS1138-262). 

The fields of MRC0114-211 ($z=1.41$), 3C470 ($z=1.65$), MRC2224-273 ($z=1.68$), 
MRC0350-279 ($z=1.90$), MRC2139-292 ($z=2.55$) and MRC2025-218 ($z=2.63$) have 
never been studied for overdensities in the past. They show several clumps of red sources 
whose association with the targeted radio galaxy still need to be confirmed with spectroscopy. 
We next briefly discuss each of these cluster candidates in order of
increasing redshift.

\subsection{MRC0114-211 ($z=1.41$)}

The environment of this radio galaxy had never been studied in the past. This field is 
the lowest redshift but also the strongest galaxy cluster candidate in the HzRG sample, 
overdense at a $6\sigma$ level compared to SWIRE. Only $0.03$\% of the cells in our 
counts-in-cells analysis are found with such density. The distribution of the IRAC-selected 
red sources shows a clear spatial segregation. The bright red galaxies lie preferentially 
in two N/S elongated structures with the radio galaxy 
embedded in the southern one. The northern clump is $2\arcmin$ ($\sim1$~Mpc) 
north of the radio galaxy. Additional groups of red galaxies 
are also observed within $1$~Mpc of the radio position, suggesting that MRC0114-211 
is part of a complex large-scale structure. Furthermore, five red IRAC-selected sources 
(including the radio galaxy itself) are found in the immediate vicinity (within 
$15\arcsec\sim125$~kpc) of the HzRG, corresponding to a density about $9$ times higher 
than average. No other radio galaxy in our sample shows such a dense environment.

\subsection{7C1756+6520 ($z=1.42$)}

\citet{Galametz2009B} identified this HzRG as residing in a candidate high-redshift 
cluster based on $BzK$ photometry. This cluster was recently confirmed in
spectroscopy with $21$ cluster members \citep{Galametz2010a}.
The spatial distribution of the IRAC-selected sources is clumpy, with a main compact clump 
of bright galaxies $1\arcmin$ SE of the radio galaxy. This location coincides with the 
spectroscopically confirmed sub-group of galaxies in \citet{Galametz2010a} at $z=1.437$. 
Seven of the spectroscopically confirmed members were observed and detected both at $3.6$ and 
$4.5\mu$m. Only one \citep[ID Cl~1756.20 in][]{Galametz2010a} 
has an unexpectedly blue IRAC color ($[3.6]-[4.5]=-0.4$). The other six galaxies have $[3.6]-[4.5]>-0.1$. 

\subsection{3C470 ($z=1.65$) and MRC2224-273 ($z=1.68$)}

These fields present very inhomogenous distributions of IRAC-selected sources. In 
both cases, the radio galaxies are located (in projection) at the edges of the galaxy 
overdensities. Spectroscopic follow-up of the structures is required 
to confirm the association of high-redshift galaxy structures with the radio galaxies. 

\subsection{MRC0350-279 ($z=1.90$)}

This is the third densest field in our sample. Two very compact clumps 
of red sources are found within $1\arcmin$ and aligned with MRC0350-279 
in the direction NW/SE, forming an intriguing symmetrical structure.

\subsection{MRC0156-252 ($z=2.02$)} 

\citet{Galametz2010B} presented a study of the surroundings of MRC0156-252. In 
that work, we used a purely near-infrared color criterion to select $z>2$ galaxies 
in the field and found that the radio galaxy lies in an overdense region of both early-type 
and star-forming candidates. The majority of the IRAC-selected sources within $1\arcmin$ 
of MRC0156-252 are found north of the radio galaxy, a result also found for the near-infrared 
selected cluster candidates in \citet{Galametz2010B}. 

\subsection{PKS1138-262 ($z=2.16$)}

This radio galaxy (also referred to as the ``Spider Web'' galaxy) is associated with a well 
studied high-redshift galaxy protocluster. Several intense programs have studied this field,
identifying a diversity of cluster members: Ly$\alpha$ emitters, 
H$\alpha$ emitters \citep{Pentericci2000, Kurk2004B}, or candidate cluster members: EROs 
\citep{Kurk2004A}, Distant Red Galaxies \citep[DRGs;][]{Kodama2007}.
To date, only two massive red sources have been spectroscopically confirmed to
be associated with PKS1138-262 \citep{Doherty2010}; both are 
red IRAC-selected sources with $[3.6]-[4.5]>0.1$.
Similar to DRGs selected in this field, the (brightest) IRAC-selected galaxies 
are preferentially found along a $3\arcmin$-long E/W filamentary structure 
(corresponding to the radio axis) in which the radio galaxy is embedded.



\subsection{USS1425-148 ($z=2.35$)}

\begin{figure*}
\begin{center} 
\includegraphics[width=5.5cm]{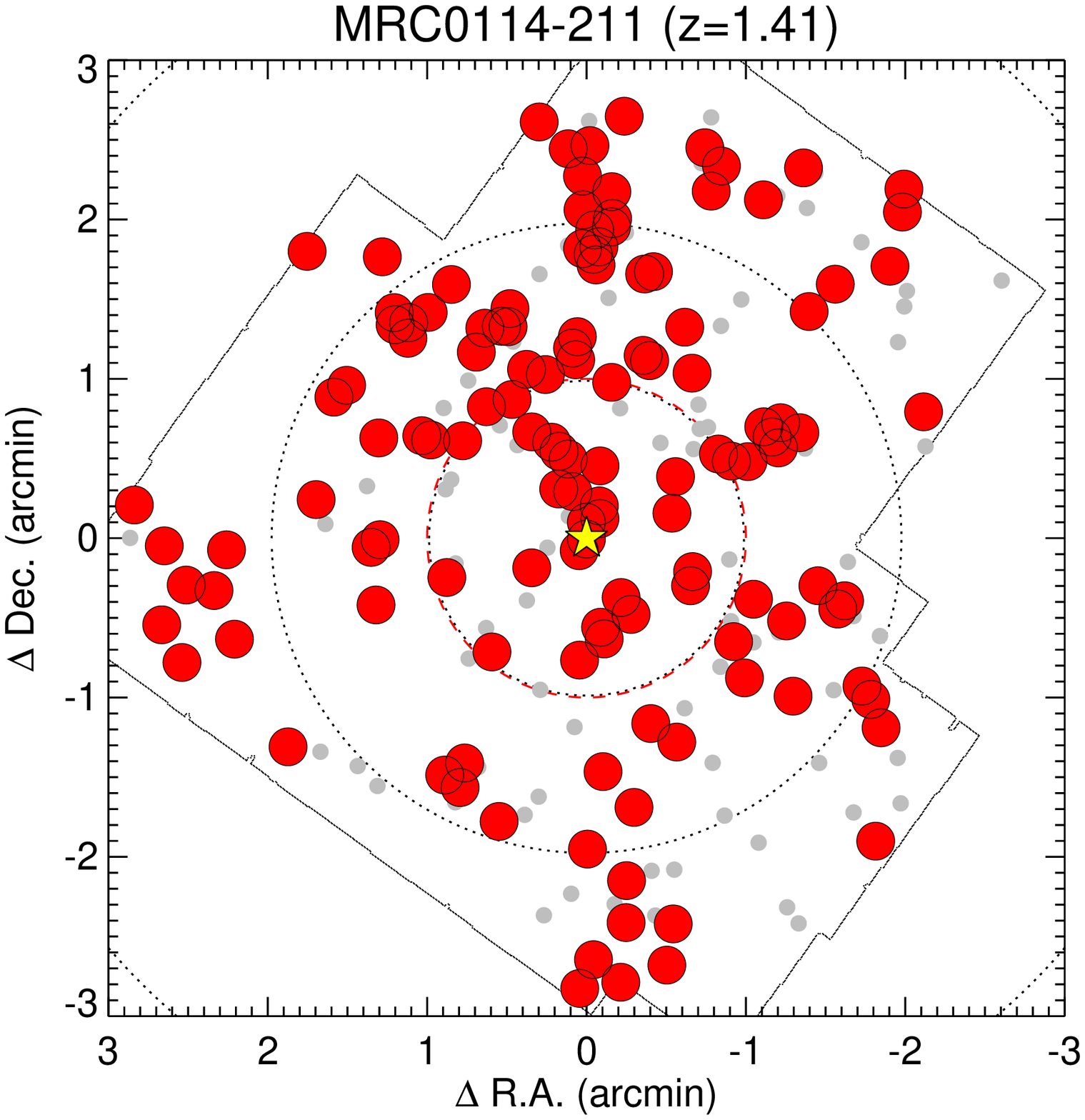}
\includegraphics[width=5.5cm]{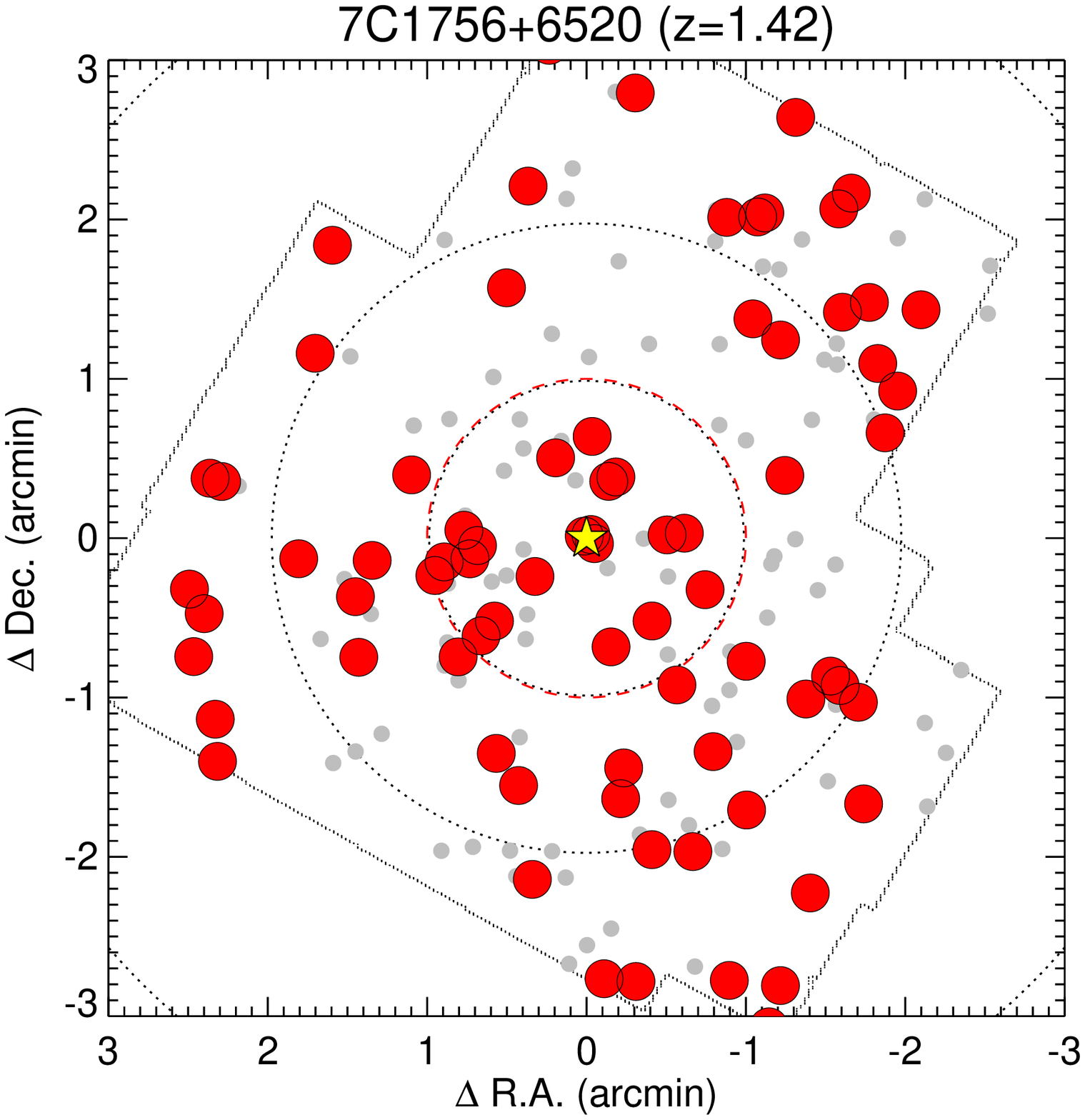}
\includegraphics[width=5.5cm]{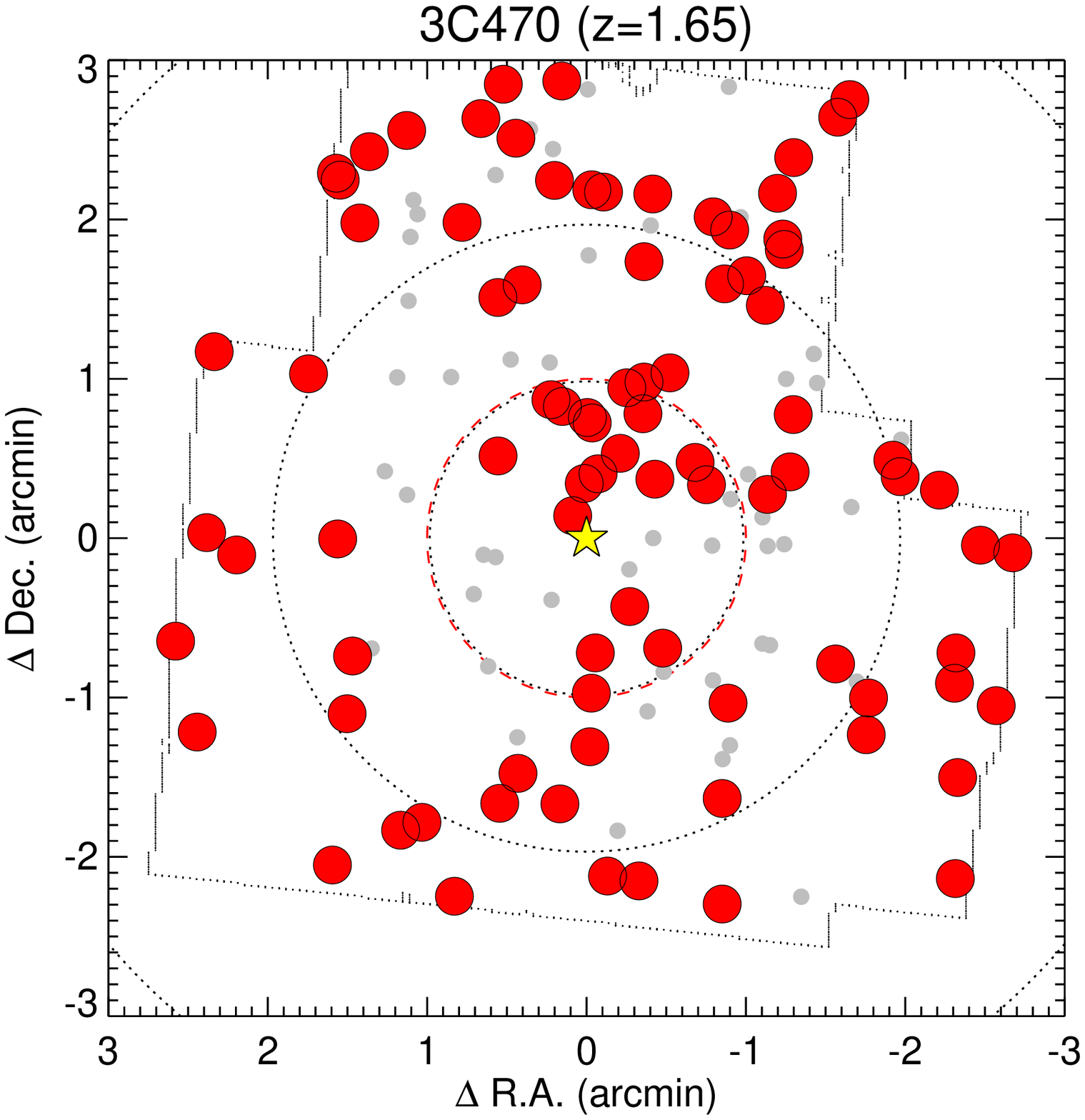}\\
\includegraphics[width=5.5cm]{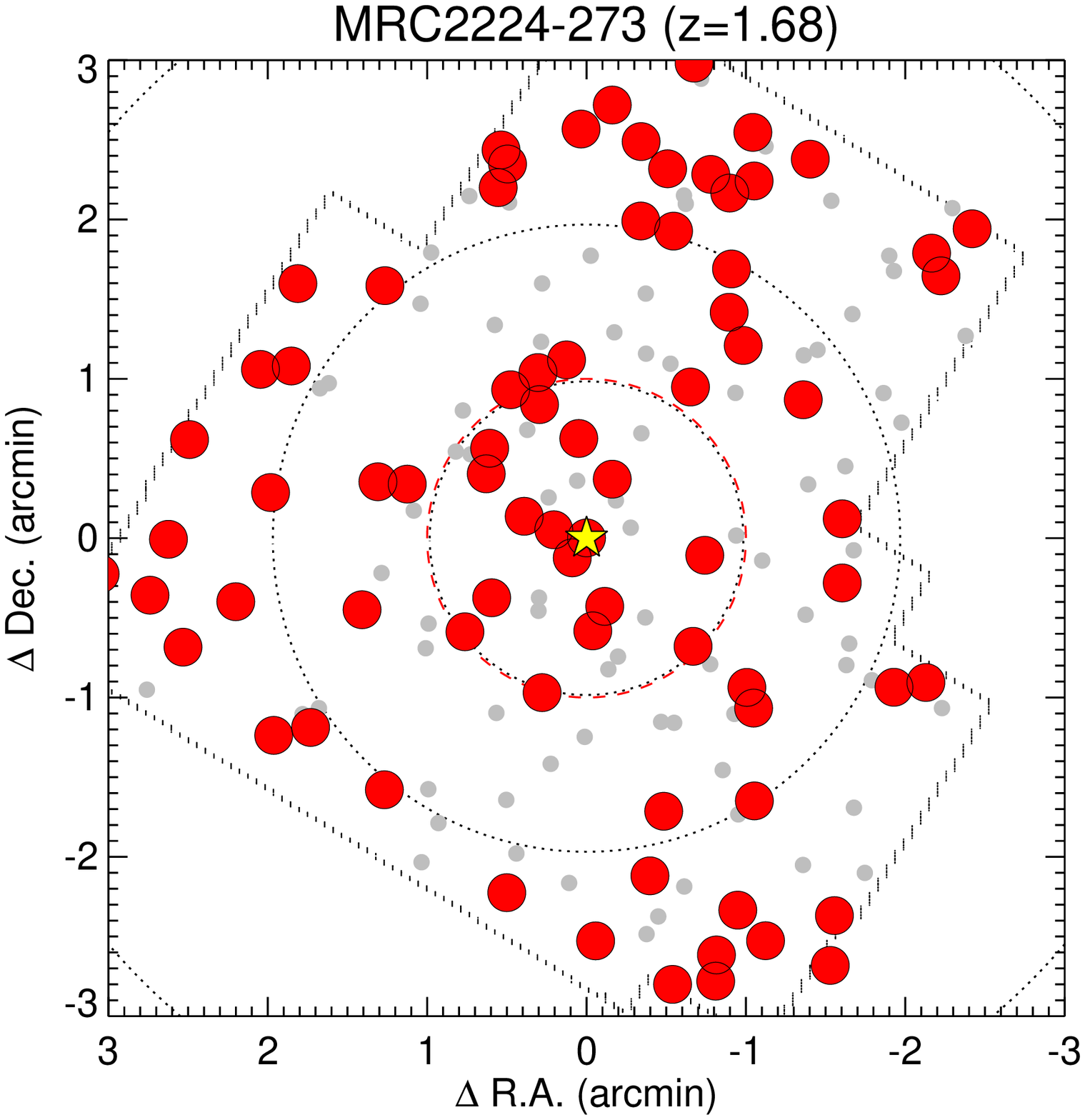}
\includegraphics[width=5.5cm]{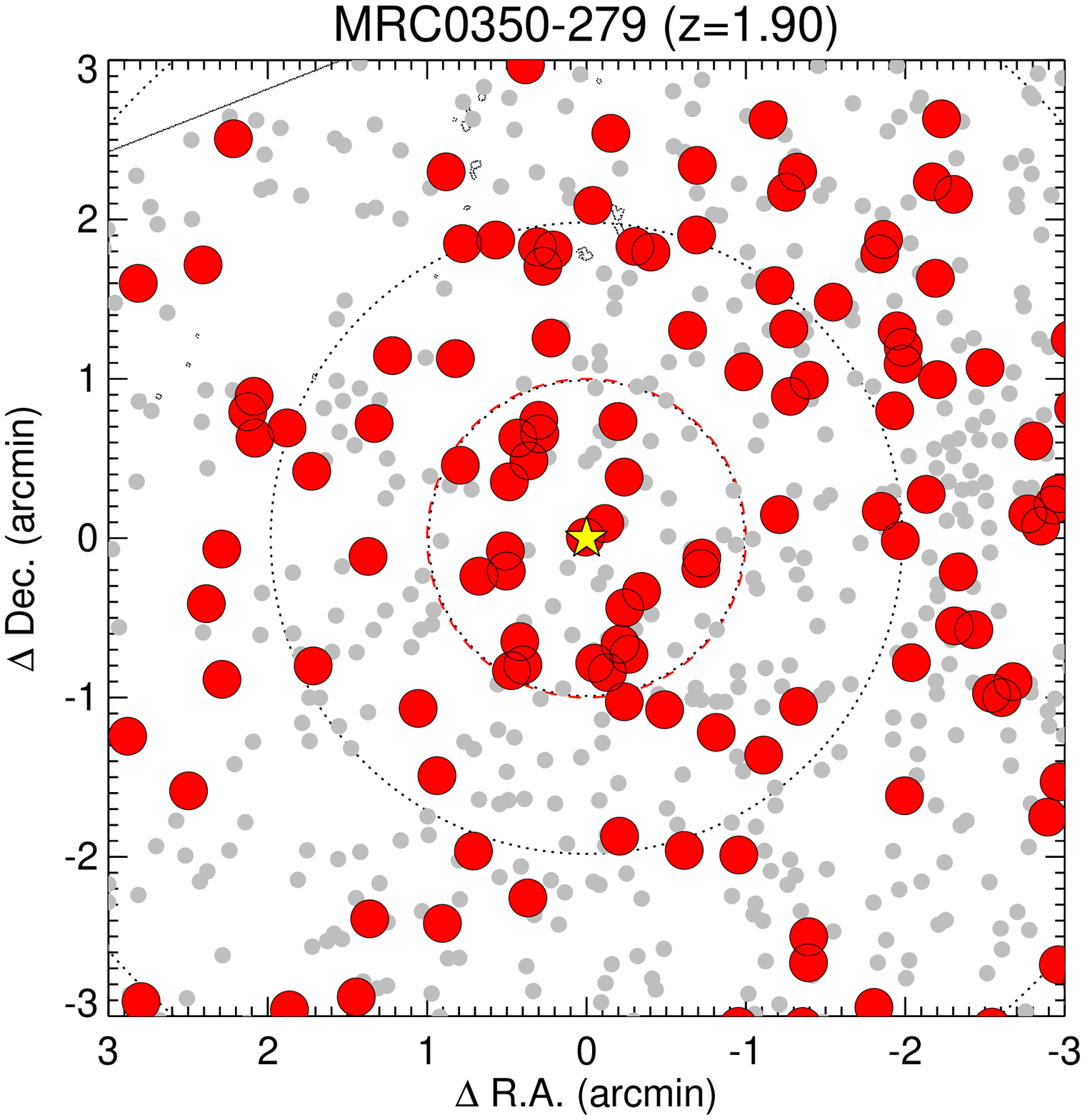}
\includegraphics[width=5.5cm]{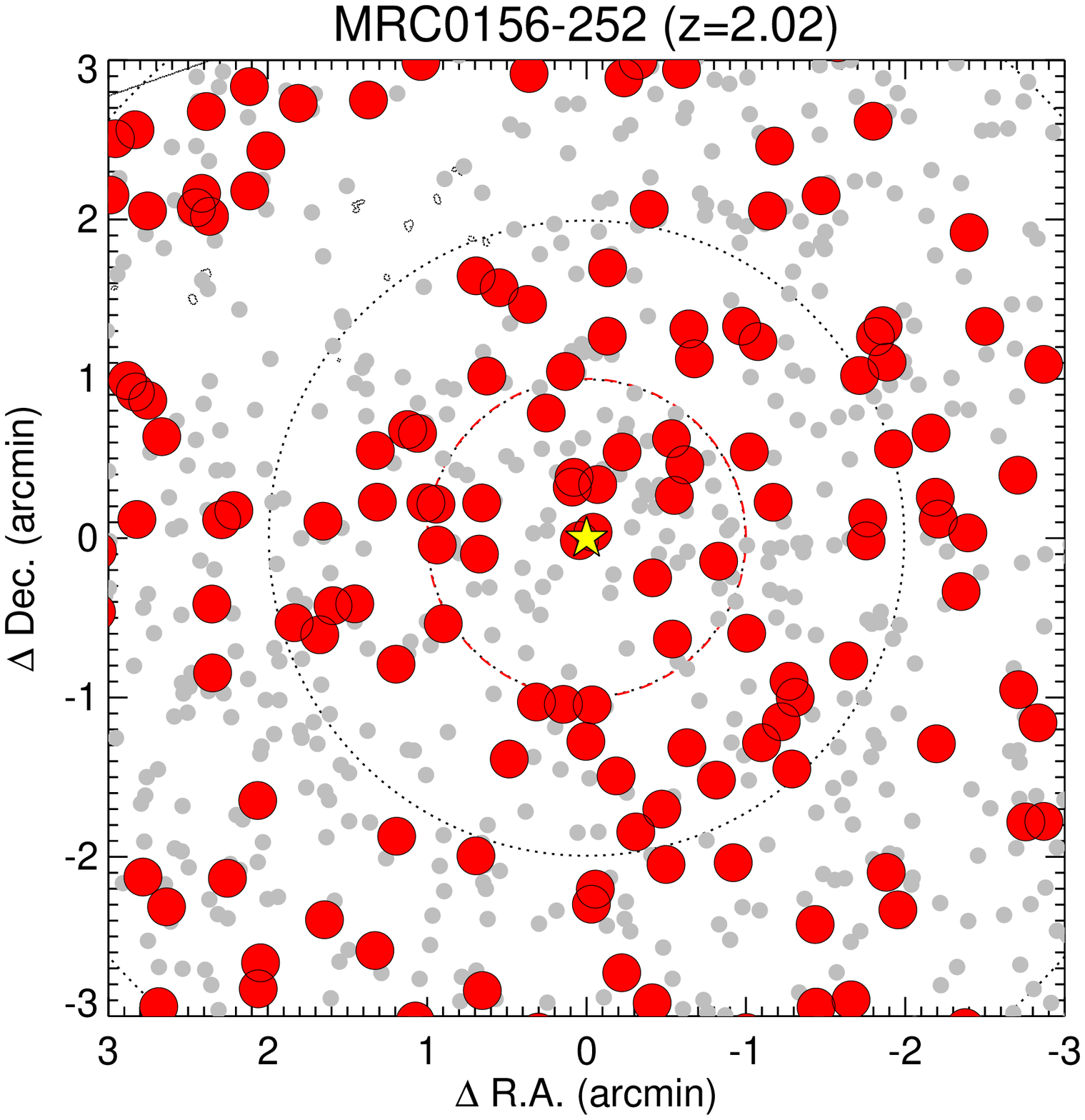}\\
\includegraphics[width=5.5cm]{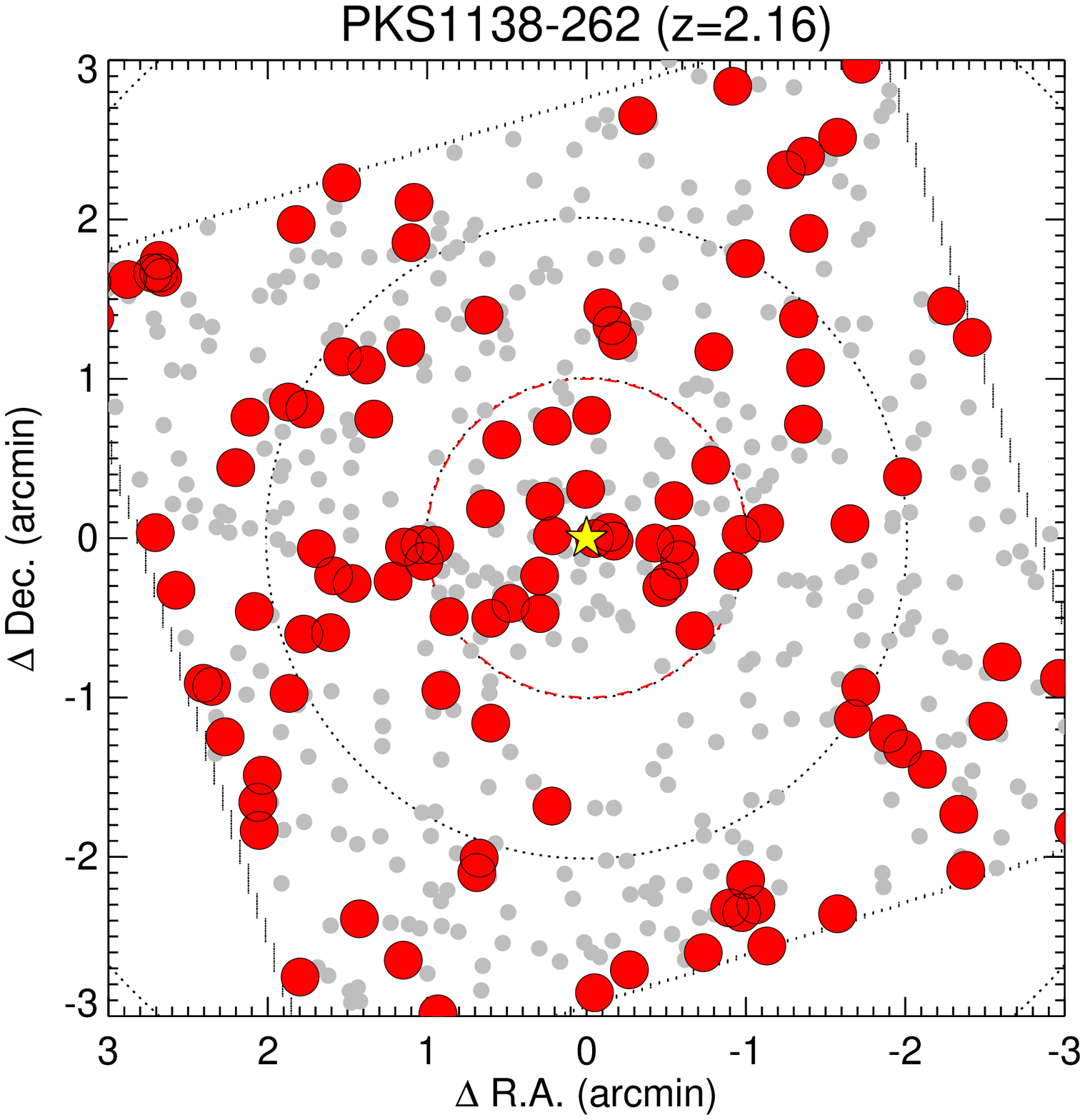}
\includegraphics[width=5.5cm]{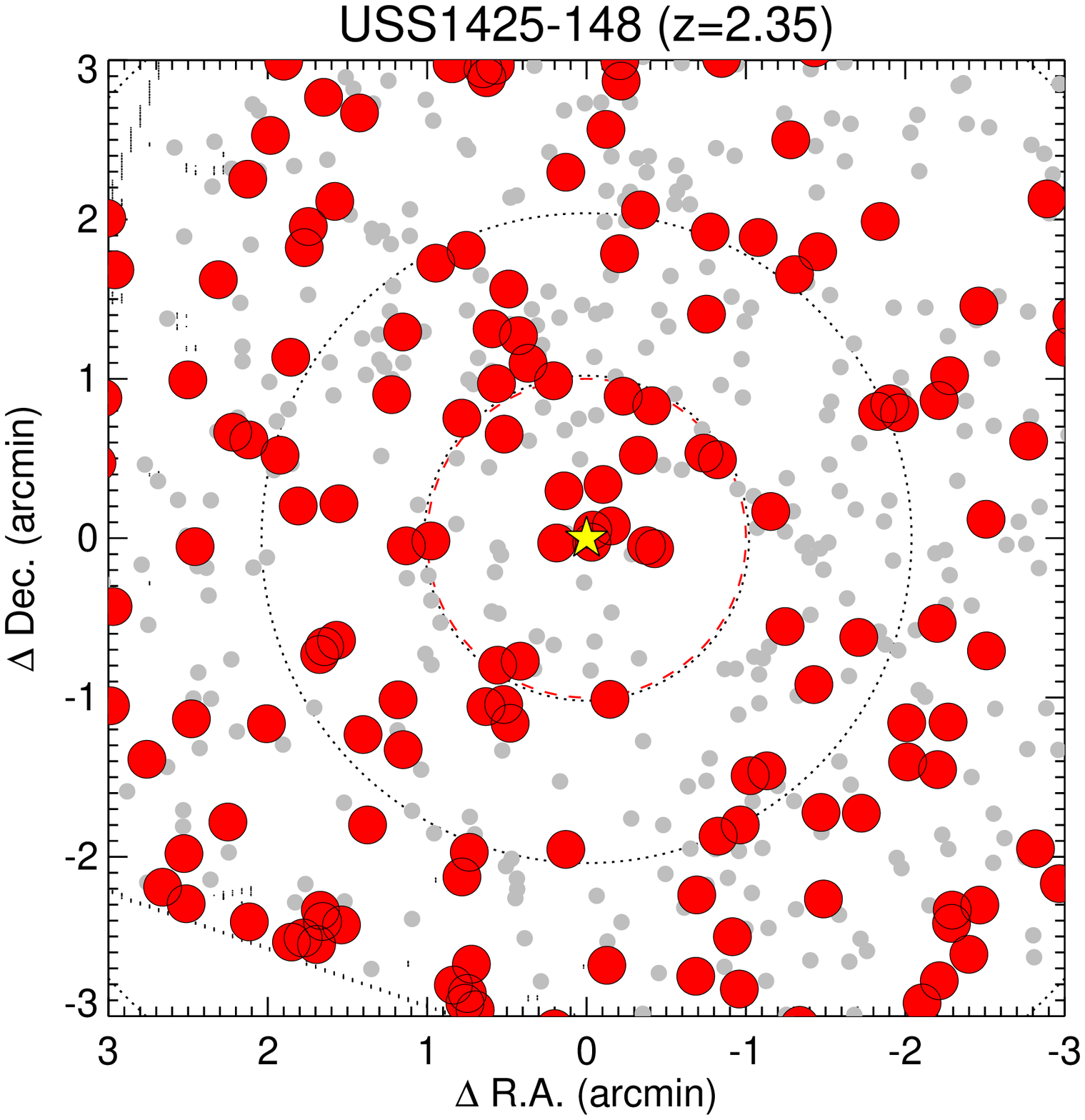}
\includegraphics[width=5.5cm]{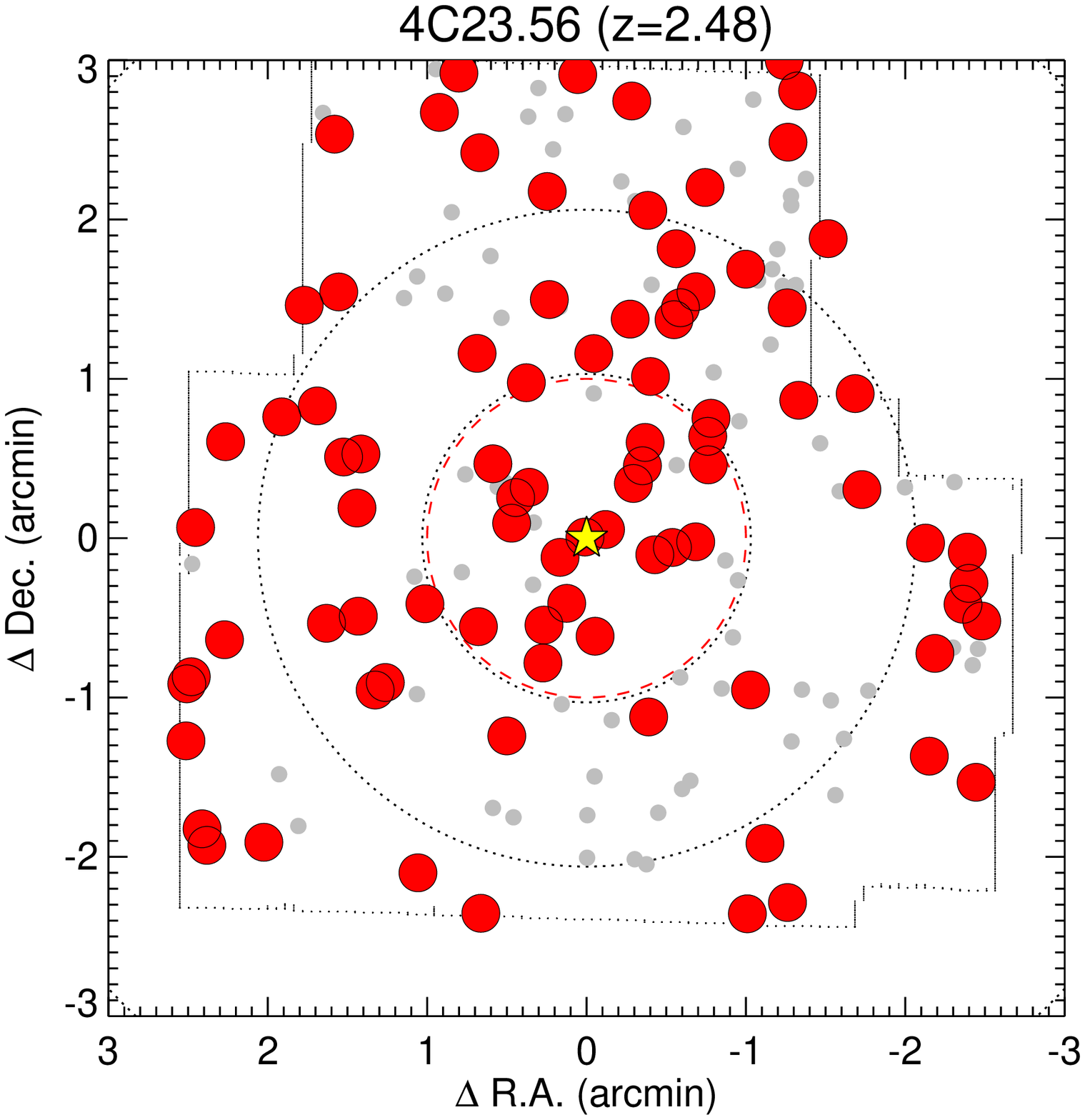}\\
\includegraphics[width=5.5cm]{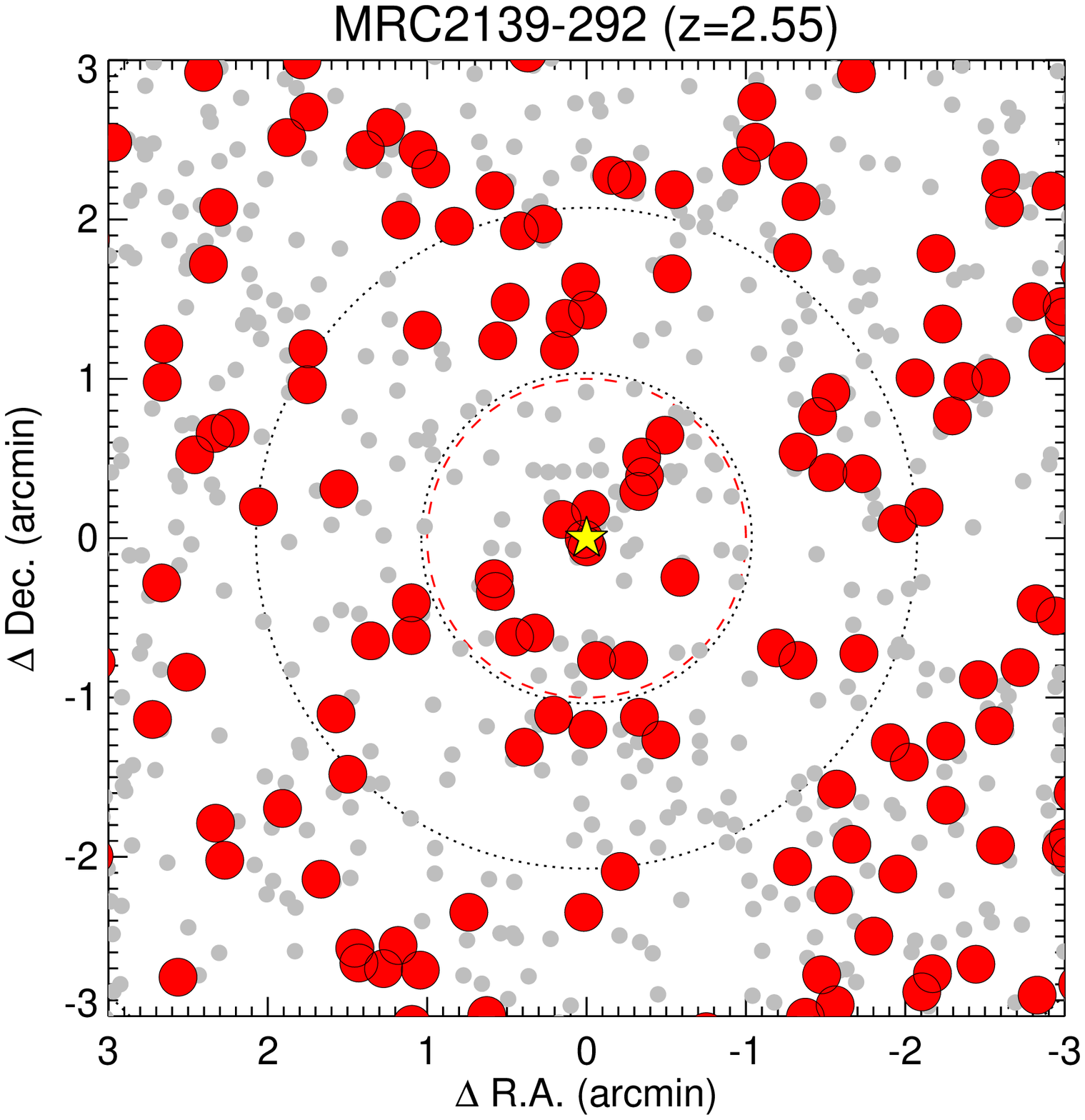}
\includegraphics[width=5.5cm]{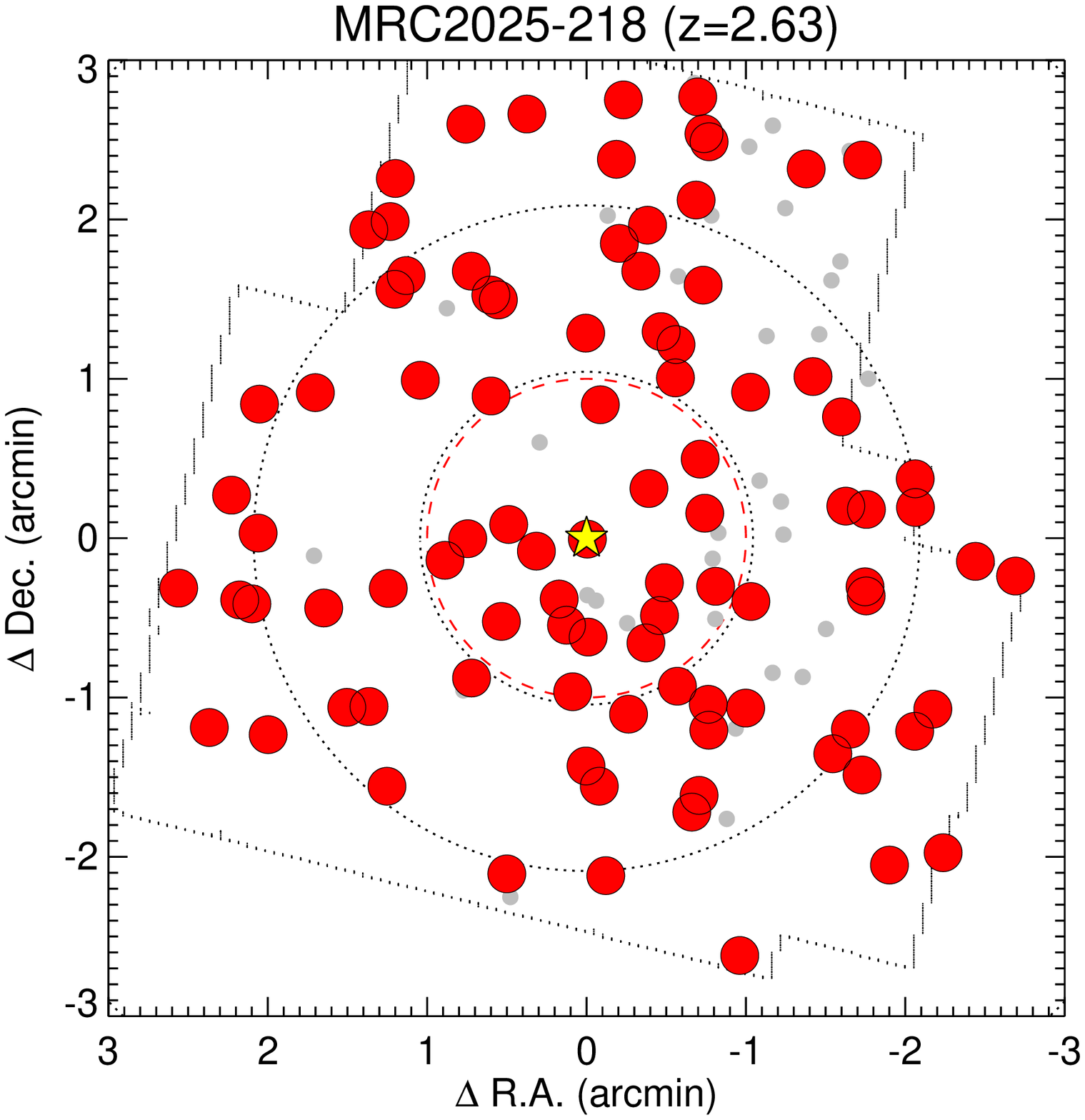}\\
\end{center}
\caption{Spatial distribution of $[3.6]-[4.5]>-0.1$ galaxies around the radio galaxies 
with a $2\sigma$ excess of IRAC-selected sources compared to SWIRE (i.e.,~with more 
than $15$ IRAC-selected sources within $1\arcmin$). 
We plot sources down to the $3\sigma$ limits of each image (small gray dots) and 
to the uniform ($5\sigma$) limits of our full sample (larger red dots). Contours of the combined 
IRAC1+2 studied zone are illustrated by the dashed lines. Concentric dotted circles account 
for distances of $0.5$, $1$ and $2$~Mpc at the redshift of the targeted radio galaxy (yellow 
star). The red dashed circle shows the cell of $1\arcmin$ radius centered on the 
HzRG and used in the count-in-cell analysis in Section 5. North is up, East is to the left.}
\label{radec} 
\end{figure*}

The environment of this radio galaxy was recently studied using near-infrared observations. 
\citet{Hatch2011} selected high-redshift galaxies using near-infrared color criteria in the field 
of six radio galaxies at $z\sim2.4$. USS1425-148 was the most overdense field of their sample
and suspected to contain a high-redshift galaxy cluster at $z>2$. Fig.~\ref{radec} suggests that 
the radio galaxy has several nearby companions. A second clump of sources is found $\sim1\arcmin$
north of USS1425-148.

We note that MRC1324-262 ($z=2.28$) and MRC0406-244 ($z=2.427$) were also part of the 
\citet{Hatch2011} sample and were not found to lie in particularly overdense environments, a 
result consistent with the present mid-infrared analysis of these fields.

\subsection{4C23.56 ($z=2.48$)}

4C23.56 is also known to lie in an overdense region of galaxies. More than a decade ago, 
\citet{Knopp1997} found an overdensity of red galaxies ($I-K>4$, Vega) in the immediate 
surroundings of 4C23.56. A complementary analysis was made by \citet{Kajisawa2006} 
who found that 4C23.56 was the densest field out of their sample of six radio galaxies at 
$2.3<z<2.6$ in terms of near-infrared cluster galaxy candidates. An excess of candidate 
H$\alpha$ emitters has also recently been detected in the field by \citet{ITanaka2011},
which also reported on the first spectroscopically confirmed 
companions for 4C23.56, three H$\alpha$ emitters at $z=2.49$.

\subsection{MRC2139-292 ($z=2.55$) and MRC2025-218 ($z=2.63$)}

These fields are our two highest redshift candidates. The field of MRC2139-292 presents a 
very clear and compact concentration of red galaxies centered on the radio galaxy. Though 
MRC2025-218 is just above our threshhold for identifying rich IRAC environments, we note 
that Mayo et al.~2012 (in press) find that it resides in a very rich field in the MIPS $24\mu$m bandpass, 
overdense at $\sim 7 \sigma$ level compared to control fields.

\section{The mid-infrared environments of $z>3$ radio galaxies}

The initial sample of HzRGs also comprised $21$ galaxies at $z>3$ that were analyzed 
in an identical manner to the main sample. Some of these radio galaxies are known to be 
within spectroscopically confirmed high-redshift protoclusters (e.g.~USS0943-242, TNJ1338-1942, 
MRC0316-257, TNJ0924-2201; Venemans et al.~2007) though none were recovered by the 
present mid-infrared analysis. This suggests evolution in the luminosity function of high 
redshift galaxy clusters. As seen in Fig.~2, the simplest single stellar population models 
predict relatively constant [4.5] flux densities at $z > 1$ for a wide range of formation redshifts, 
as well as a brightening of the galaxies as we approach the formation epoch. The fact that we 
do not recover these known rich (proto-)clusters at $z > 3$ implies that this simple model is 
incorrect, most likely because the massive ellipticals that comprise the clusters are still in 
process of forming in a hierarchical manner. Deeper IRAC data from a large sample of 
clusters and cluster candidates out to $z\sim 3$ should test this hypothesis. We are, in fact, 
amassing such a data set with the Clusters Around Radio-Loud AGN (CARLA)
{\it Warm Spitzer} snaphot program (Galametz et al. in prep.).

The radio galaxy 4C41.17 ($z=3.79$) lies in an overdense region of IRAC-selected sources. A compact
clump of IRAC-selected sources is observed, slightly offset $1\arcmin$ south of the radio galaxy. 
\citet{Ivison2000} also reported an overdensity of sub-millimeter galaxies in this field, though spectroscopic 
follow-up has failed to confirm their association with the radio galaxy \citep{Greve2007}. We suspect 
that there might be a foreground galaxy structure in this field.

\section{Conclusions}

We identified overdensities in the fields of powerful HzRGs at $z>1.2$. We 
studied the distribution of sources with $[3.6]-[4.5]>-0.1$ using a counts-in-cell analysis,
defining a region to be overdense when denser (at the $2\sigma$ level) 
than the SWIRE average, i.e.,~when $15$ or more red IRAC-selected galaxies 
(with $f_{4.5}\geq13.4\mu$Jy) are found in a cell of $1\arcmin$ radius. 
We identified $11$ radio galaxies with at least $15$ IRAC-selected candidates within 
$1\arcmin$. 
These $11$ radio galaxy fields, representing $23$\% of our sample of $48$ HzRGs, are 
the most promising galaxy cluster candidates from our analysis. The large fraction of 
overdense fields clearly indicates that radio galaxies, on average, reside in rich enviroments. 
Randomly selecting similarly analyzed samples of $48$ control fields from the SWIRE survey, 
we find such a high fraction of dense fields in only $0.3$\% of the samples.

Five of these fields have been studied in the past and are known or suspected to contain a 
(proto-)cluster at high-redshift. Three have been spectroscopically confirmed 
to be associated with the radio galaxy. This clearly demonstrates the strength 
of IRAC to select high-redshift galaxy structures. Our relatively shallow data have 
permitted the recovery of known clusters in very short integration times ($120$s); 
past ground-based studies have required hours of optical/near-infrared observations. 

We tested for correlations of environment with radio galaxy properties and found no
significant trends with either redshift or radio luminosity. We note however that a detailed 
analysis of evolution of the cluster population with redshift will require 
an even larger sample of high-redshift clusters as well as a better understanding of 
the luminosity function for clusters at $z>1$. No statistically significant trend with radio 
luminosity was observed. Nevertheless, the majority of our cluster candidates are found 
in the fields of more luminous radio galaxies ($L_{\rm 500MHz}>10^{28.6}$~W Hz$^{-1}$). 

\begin{acknowledgements}
This work is based on observations made with the {\it Spitzer Space Telescope}, 
which is operated by the Jet Propulsion Laboratory, California Institute of 
Technology under a contract with NASA. We are very grateful to Mark Brodwin 
and Peter Eisenhardt for having provided information on the Bo\"{o}tes cluster sample
mentioned in this paper and to Conor Mancone for providing his useful EZ Gal Model
Generator and valuable help on models. We also thank the anonymous referee for his/her
very useful comments.

\end{acknowledgements}

\clearpage
\clearpage

\begin{table*}
\caption{The shallow HzRG sample (in redshift order).}
\label{targets}
\centering
\begin{tabular}{l c c c c c}
\hline
Name	&	R.A.			&	Dec.			&	{\it z}$^{\mathrm{a}}$		&	log($L_{\rm 500MHz}$)$^{\mathrm{b}}$	&	Density	\\
		&	(J2000)		&	(J2000)		&			&	(W Hz$^{-1}$)						&	(arcmin$^{-2}$)	\\
\hline
\hline
3C356.0	        		&	17:24:19.0	&	 50:57:40.30	&	  1.079	&	28.35	&     $2.5\pm0.9$	\\	
MRC0037$-$258	&	 00:39:56.4	&	$-$25:34:31.01	&	  1.100	&	27.72  	&     $3.5\pm1.1$	\\	
3C368.0	        		&	18:05:06.3	&	 11:01:33.00	&	  1.132	&	28.52	&     $3.2\pm1.0$ 	\\	
6C0058+495		&	 01:01:18.9	&	 49:50:12.29	&	  1.173	&	27.33	&     $3.8\pm1.1$	\\
\hline	
3C266			&	11:45:43.4	&	 49:46:08.24	&	  1.275	&	28.54     	&     $3.8\pm1.1$	\\	
MRC0211$-$256	&	 02:13:30.5	&	$-$25:25:21.00	&	  1.300	&	27.78	&     $2.5\pm0.9$	\\		
MRC0114$-$211	&	 01:16:51.4	&	$-$20:52:06.71	&	  1.410	&	28.66	&     $8.9\pm1.7$	\\	
7C1756+6520		&	17:57:05.4	&	 65:19:53.11	&	  1.4156	&	27.40	&     $6.4\pm1.4$	\\	
7C1751+6809		&	17:50:49.9	&	 68:08:25.93	&	  1.540	&	27.46	&     $2.5\pm0.9$	\\	
3C470			&	23:58:35.3	&	 44:04:38.87	&	  1.653	&	28.79	&     $5.7\pm1.4$	\\	
MRC2224$-$273	&	22:27:43.3	&	$-$27:05:01.71	&	  1.679       &	27.52  	&     $4.8\pm1.2$	\\	
6C0132+330		&	 01:35:30.4	&	 33:17:00.82	&	  1.710	&	27.64	&     $4.1\pm1.1$	\\	
3C239			&	10:11:45.4	&	 46:28:19.75	&	  1.781	&	29.00	&     $2.2\pm0.8$	\\	
3C294.0			&	14:06:44.0	&	 34:11:25.00	&	  1.786       &	28.96  	&     $2.9\pm1.0$	\\	
7C1805+6332		&	18:05:56.9	&	 63:33:13.14	&	  1.840	&	27.78	&     $4.5\pm1.2$	\\	
6CE0820+3642	&	08:23:48.1	&	 36:32:46.42	&	  1.860	&	28.28	&     $1.6\pm0.7$	\\	
6CE0905+3955	&	09:08:16.9	&	 39:43:26.00	&	  1.883	&	28.17	&     $3.5\pm1.0$	\\	
6CE0901+3551	&	09:04:32.4	&	 35:39:03.23	&	  1.910	&	28.19	&     $3.2\pm1.0$	\\	
MRC0152$-$209	&	 01:54:55.8	&	$-$20:40:26.30	&	  1.920	&	28.20	&     $3.8\pm1.1$	\\	
MRC2048$-$272	&	20:51:03.6	&	$-$27:03:02.53	&	  2.060	&	28.72	&     $4.5\pm1.2$	\\	
5C7.269			&	08:28:38.8	&	 25:28:27.10	&	  2.218	&	27.82	&     $2.2\pm0.8$	\\	
4C40.36			&	18:10:55.7	&	 40:45:24.01	&	  2.265     	&	28.79	&     $2.9\pm1.0$	\\	
TXS0211$-$122	&	 02:14:17.4	&	$-$11:58:46.00	&	  2.340	&	28.48	&     $4.1\pm1.1$	\\		
USS1707+105		&	17:10:06.5	&	 10:31:06.00	&	  2.349       &	28.63   	&     $1.6\pm0.7$	\\	
USS1410$-$001	&	14:13:15.1	&	 $-$00:22:59.70	&	  2.363	&	28.41	&     $1.3\pm0.6$	\\	
6C0930+389		&	09:33:06.9	&	 38:41:50.14	&	  2.395	&	28.41	&     $2.2\pm0.8$	\\	
3C257			&	11:23:09.2	&	  05:30:19.47	&	  2.474	&	29.16	&     $3.2\pm1.0$	\\	
4C23.56			&	21:07:14.8	&	 23:31:45.00	&	  2.483       &	28.93  	&     $6.4\pm1.4$	\\
USS1558$-$003	&	16:01:17.3	&	 $-$00:28:48.00	&	  2.527       &	28.82	&	$3.8\pm1.1$      \\
WNJ1115+5016	&	11:15:06.9	&	 50:16:23.92	&	  2.540	&	27.82	&     $1.3\pm0.6$	\\
USS0828+193		&	08:30:53.4	&	 19:13:16.00	&	  2.572       &	28.44  	&     $3.2\pm1.0$	\\
PKS0529$-$549	&	 05:30:25.2	&	$-$54:54:22.00	&	  2.575	&	29.16	&     $2.2\pm0.8$	\\
MRC2025$-$218	&	20:27:59.5	&	$-$21:40:56.90	&	  2.630	&	28.74	&     $5.7\pm1.3$	\\ 
USS2202+128		&	22:05:14.1	&	 13:05:33.50	&	  2.706	&	28.54	&     $2.2\pm0.8$	\\
MG1019+0534		&	10:19:33.4	&	  05:34:34.80	&	  2.765	&	28.57	&     $3.2\pm1.0$	\\
4C24.28			&	13:48:14.8	&	 24:15:52.00	&	  2.879	&	29.05	&     $2.9\pm1.0$	\\
4C28.58			&	23:51:59.2	&	 29:10:28.99	&	  2.891       &	28.91   	&     $3.8\pm1.1$	\\
USS0943$-$242	&	09:45:32.7	&	$-$24:28:49.65	&	  2.923	&	28.62	&     $2.9\pm1.0$	\\
WNJ0747+3654	&	07:47:29.4	&	 36:54:38.09	&	  2.992	&	28.14	&     $2.5\pm0.9$	\\
\hline
B3J2330+3927		&	23:30:24.9	&	 39:27:12.02	&	  3.086	&	28.33	&     $1.3\pm0.6$	\\
WNJ0617+5012	&	 06:17:39.4	&	 50:12:55.40	&	  3.153	&	28.02	&     $3.9\pm1.1$	\\
MRC0251$-$273	&	 02:53:16.7	&	$-$27:09:13.03	&	  3.160      &	28.54   	&     $2.5\pm0.9$	\\
WNJ1123+3141	&	11:23:55.9	&	 31:41:26.14	&	  3.217	&	28.51	&     $2.9\pm1.0$	\\
6C1232+39		&	12:35:04.8	&	 39:25:38.91	&	  3.220      &	28.93   	&     $2.5\pm0.9$	\\
TNJ0205+2242	&	 02:05:10.7	&	 22:42:50.40	&	  3.506      &	28.46    	&     $1.6\pm0.7$	\\
TNJ0121+1320	&	 01:21:42.7	&	 13:20:58.00	&	  3.516	&	28.49	&     $3.2\pm1.0$	\\
TXJ1908+7220		&	19:08:23.7	&	 72:20:11.82	&	  3.53         &	29.12 	&     $5.4\pm1.3$	\\
USS1243+036		&	12:45:38.4	&	  03:23:20.70	&	  3.570	&	29.23	&     $4.1\pm1.1$	\\
WNJ1911+6342	&	19:11:49.6	&	 63:42:09.60	&	  3.590	&	28.14	&     $1.9\pm0.8$	\\
MG2144+1928		&	21:44:07.5	&	 19:29:14.60	&	  3.592	&	29.08	&     $3.5\pm1.1$	\\
6C0032+412		&	  00:34:53.1	&	 41:31:31.50	&	  3.670       &	28.75	&     $1.9\pm0.8$	\\
4C60.07			&	 05:12:54.8	&	 60:30:52.01	&	  3.788	&	29.20	&     $4.1\pm1.1	$	\\
TNJ2007$-$1316	&	20:07:53.3	&	$-$13:16:43.62	&	  3.840	&	29.13	&     $3.2\pm1.0$	\\
8C1435+635		&	14:36:37.1	&	 63:19:14.00	&	  4.250	&	29.40	&     $2.5\pm0.9$	\\
TNJ0924$-$2201	&	09:24:19.9	&	$-$22:01:41.00	&	  5.195	&	29.51	&     $1.0\pm0.6$	\\	
\hline     
\end{tabular}       
\begin{list}{}{}
\item[$^{\mathrm{a}}$] Our analysis focusses on fields around HzRGs with $1.2<z<3$. The other fields are used 
as control fields. 
\item[$^{\mathrm{b}}$] Radio luminosities from \citet{DeBreuck2010}.
\end{list}
\end{table*}

\begin{table*}
\caption{The deeper Spitzer sample (in redshift order).}
\label{targets2}
\centering
\begin{tabular}{l c c c c c c c}
\hline
Name			&	R.A.			&	Dec.			&	{\it z}	$^{\mathrm{a}}$	&	log($L_{\rm 500MHz}$)$^{\mathrm{a}}$	&	IRAC Exp. Time	&	PID	&	Density	\\
				&	(J2000)		&	(J2000)		&			&	(W Hz$^{-1}$)			&		(s)		&		&	(arcmin$^{-2}$)	\\
\hline
\hline	
LBDS53W091			&	17:22:32.7	&	 50:06:01.94	&	  1.552       &	27.04   	&	900		&	65		&	$3.8\pm1.1$	\\
MRC1017$-$220		&	10:19:49.0	&	$-$22:19:58.03	&	  1.768       &	27.94   	&	1600		&	60112	&	$4.5\pm1.2$	\\
MRC0324$-$228		&	 03:27:04.4	&	$-$22:39:42.60	&	  1.894       &	28.49   	&	1600		&	60112	&	$2.9\pm1.0$	\\
MRC0350$-$279		&	 03:52:51.6	&	$-$27:49:22.61	&	  1.900	&	28.25	&	1600		&	60112	&	$8.0\pm1.6$	\\
MRC0156$-$252		&	 01:58:33.6	&	$-$24:59:31.10	&	  2.016       &	28.46   	&	1600		&	60112	&	$5.1\pm1.3$	\\
PKS1138$-$262		&	11:40:48.6	&	$-$26:29:08.50	&	  2.156	&	29.07	&	3000		&	17		&	$8.3\pm1.6$	\\
MRC1324$-$262		&	13:26:50.82	&	$-$26:30:51.10	&	  2.280	&	28.45	&	1600		&	60112	&	$1.9\pm0.8$	\\
USS1425$-$148		&	14:28:52.50	&	$-$15:01:37.50	&	  2.349	&	28.66	&	1600		&	60112	&	$5.4\pm1.3$	\\
LBDS53W002			&	17:14:14.7	&	 50:15:29.70	&	  2.393	&	27.78	&	3300		&	211		&	$4.1\pm1.1$	\\
MRC0406$-$244		&	 04:08:51.5	&	$-$24:18:16.39	&	  2.427       &	29.03   	&	1600		&	60112	&	$3.8\pm1.1$	\\
MG2308+0336			&	23:08:28.05	&	03:36:20.70	&	  2.457 	&	28.51	&	1600		&	60112	&	$4.5\pm1.2$	\\
MRC2104$-$242		&	21:06:58.1	&	$-$24:05:11.00	&	  2.491	&	28.84	&	1600		&	60112	&	$4.1\pm1.1$	\\
MRC2139$-$292		&	21:42:16.7	&	$-$28:58:40.00	&	  2.550	&	28.73	&	1600		&	60112	&	$4.8\pm1.2$	\\
\hline	
MRC0316$-$257		&	 03:18:12.0	&	$-$25:35:11.00	&	  3.130      &	28.95    	&	46000	&	3482		&	$3.8\pm1.1$	\\
B20902+34			&	09:05:30.1	&	 34:07:56.89	&	  3.395	&	28.78	&	1200		&	64		&	$3.2\pm1.0$	\\
4C41.17				&	06:50:52.1	&	 41:30:31.00	&	  3.792       &	29.18   	&	5000		&	79		&	$6.7\pm1.5$	\\
TNJ1338$-$1942		&	13:38:26.0	&	$-$19:42:31.00	&	  4.110    	&	28.71	&	5000		&	17		&	$3.8\pm1.1$	\\
6C0140+326			&	 01:43:43.8	&	 32:53:49.31	&	  4.413       &	28.73  	&	5000		&	79		&	$4.5\pm1.2$	\\
\hline     
\end{tabular}   
\begin{list}{}{}
\item[$^{\mathrm{a}}$] Our analysis focusses on fields around HzRGs with $1.2<z<3$. The other fields are used as control fields. 
\item[$^{\mathrm{b}}$] Radio luminosities from \citet{DeBreuck2010}.
\end{list}
\end{table*}



\end{document}